\documentclass{aa}

\usepackage{epsfig}
\usepackage{multicol}
\usepackage{mdwlist}
\usepackage[switch]{lineno}
\usepackage{graphicx}
\usepackage{lscape}
\usepackage{float}
 \usepackage[varg]{txfonts}
\usepackage{longtable}
\usepackage{natbib}

\bibpunct{(}{)}{;}{a}{}{,} 
\usepackage{array}
\newcolumntype{R}{>{\raggedright\arraybackslash}p{5.7cm}}

\def\ecs{erg~s$^{-1}$cm$^{-2}$}
\def\lum{erg~s$^{-1}$}

\begin{document}

\title{BeppoSAX-WFC catalog of fast X-ray transients}

\author{J.J.M. in `t Zand\inst{1}
\and
C. Guidorzi\inst{2,3}
\and
J. Heise\inst{1}
\and
L. Amati\inst{3}
\and
E. Kuulkers\inst{4}
\and
F. Frontera\inst{2,3}
\and
G. Gianfagna\inst{5}
\and
L. Piro\inst{5}
}

\institute{SRON Space Research Organization Netherlands, Niels
Bohrweg 4, 2333 CA Leiden, the Netherlands
\and
Dept. of Physics and Earth Science, University of Ferrara, 44122 Ferrara, Italy
\and
INAF-OAS Bologna, Via P. Gobetti 101, 40129, Bologna, Italy
\and
ESTEC, ESA, Keplerlaan 1, 2201 AZ Noordwijk, the Netherlands
\and
INAF-IAPS Roma, via Fosso del Cavaliere 100, 00133 Rome, Italy
}

\date{Accepted for publication on April 22, 2026}

\abstract{ We performed a search for fast X-ray transients (FXTs),
  here defined to be transients with durations longer than one second
  and less than one day, through data of the Wide Field Camera (WFC)
  instrument onboard the BeppoSAX X-ray observatory collected between
  June 1996 and April 2002. The WFC sensitivity ranged from
  10$^{-9}$~\ecs\ (2-10 keV), for a time scale of 10 s, to a few times
  10$^{-11}$~\ecs\ for a time scale of 10$^5$ s. The WFC location
  accuracy was 0.7-4.9\arcmin\ at 68\% confidence. We focused our
  search on gamma-ray bursts (GRBs), X-ray flashes (XRFs), X-ray
  flares from high-mass X-ray binaries and stellar flares. 149 FXTs
  were detected. 63 of these are new to the literature. 38 flares are
  identified with 22 nearby stars. Three stars have never been seen
  flaring before in X-rays or optical (NLTT 51688, GR Dra and UCAC4
  255-003783). We find that the MeV transient GRO J1753+57 is most
  likely the same object as GR Dra rather than an AGN as previously
  thought. Eleven flares were detected from known high-mass X-ray
  binaries with irregular wind accretion. 100 GRBs were identified of
  which 24 have not been published before. We classify 37\% of the
  X-ray detected GRBs as XRFs with a relatively large X-ray to
  gamma-ray flux ratio, gamma-rays being measured with the BeppoSAX
  Gamma Ray Burst Monitor. The duration/spectral hardness distribution
  of all FXTs is bimodal, separating the group roughly in transients
  shorter and longer than 1 ksec and with relatively hard and soft
  spectra, respectively. We identify the 'short' FXTs as GRBs and XRFs
  and the `long' FXTs as flares from nearby late-type stars and X-ray
  binaries. The BeppoSAX-WFC FXT sample is found to be consistent with
  the one observed by Einstein Probe, when the sensitivity of the two
  instruments is taken into account, suggesting that the bulk of the
  population of faints events disclosed by Einstein Probe may
  represent the extension of a similar population to lower
  luminosities.}


\maketitle
\nolinenumbers

\section{Introduction}
\label{intro}

The X-ray sky is dynamic, more so than the visual sky. Variability is
detected on all observable time scales and up to the largest
amplitudes.  Often the dynamical range is so large that sources change
from undetectable to the brightest levels and back. If the duty cycle
of activity is low (i.e., if the flux increases for $\leq10\%$ of the
time) we call it a transient event \citep[e.g.,][]{csl97}. This
definition of a transient is somewhat arbitrary and it depends on
detector sensitivity. For our purposes, the dynamic range of the flux
variation between peak and undetectability is of order 10$^3$. The
duration of X-ray transient activity can vary between milliseconds
(e.g., soft gamma-ray repeaters and type-II bursts) to years (e.g.,
low-mass X-ray binaries). One class of transients concerns those that
last shorter than $\sim$1 day and longer than $\sim$1 s and are
generally not associated with the X-ray bright accreting compact
objects in X-ray binaries. Such transients are generally called fast
X-ray transients (FXTs) and include a mixture of different sources.

The brightest FXTs are Gamma Ray Bursts (GRBs) and stellar
flares. GRBs happen in external galaxies and are associated with two
different classes of progenitors, where the collapse of a massive star
or the merging two compact objects (neutron stars, black holes) yields
a relativistic expanding jet that produces the GRB and its afterglow
\citep[e.g.,][]{macfadyen1999,woosley2006,metzger2010}. GRBs last
shorter than 1 hr, often shorter than 1 min, and are bright in $>30$
keV $\gamma$-rays \citep[e.g.,][]{gehrels2009}. They are also bright
in 1-10 keV X-rays. In fact, it was realized that a substantial
fraction of GRBs have a soft spectrum while having similar
(statistical) properties otherwise \citep[including the spectral
  model, e.g.][]{stroh1998,heise2001,sakamoto2005}. A fair fraction of
FXTs even have no detectable $\gamma-$ray emission. These are called
X-ray flashes (XRFs). Since the detectability of $\gamma$-ray emission
depends on the sensitivity of the employed $\gamma$-ray detector, an
unbiased definition of XRFs is that the peak energy in the $\nu F_\nu$
spectrum is lower than 10 keV. However, the peak energy is often
difficult to measure and it is more practical to employ a definition
based on the ratio of measured photon flux between X-rays and
$\gamma$-rays. The ratio is typically larger than 1 for XRFs and
similar or slightly lower for the so-called X-ray rich GRB (XRR),
\citep{barraud2003,dalessio2006}. There is substantial evidence that
XRFs are simply very soft GRBs. They have the same spectral model,
timing properties and afterglows
\citep{kippen2003,soderberg2004,soderberg2006}.

Stellar flares are abundant in stars of late spectral type (i.e., that
of the Sun or later) and young T Tau stars. This was convincingly
shown by Kepler and TESS, the latter of which detected more than 120
thousand of them in the optical \citep[e.g.,][]{seli2025}. TESS
detected stellar flares in 7.5\% of all monitored stars. The X-ray
counterparts to stellar flares have peak luminosities of about
10$^{27}$-10$^{32}$~\lum\ \citep{pye2015}, except for 'superflares'
that can be up to 10$^3$ times more luminous. For typical
sensitivities of about 10$^{-10}$~\ecs, these flares can be detected
up to a few hundred parsecs distance, which renders them an isotropic
sky distribution \citep[c.f.,][]{gilmore1983} just like the
extragalactic GRBs.

FXTs have been detected ever since the advent of space-borne X-ray
astronomy. Two early reports are \cite{heise1975} and
\cite{rappaport1976}.  The first systematic studies on complete
databases of some X-ray instruments were published in the
1980s. \cite{connors1986} studies data taken by the HEAO 1 A-2
experiment in the 2 to 60 keV bandpass at a sensitivity of
$10^{-10}$~\ecs\ (on a time scale of a few seconds) and discovered
five transients with durations of 60 to 2000 s and peak fluxes up to
$10^{-9}$~\ecs. The rate corresponds to 10$^4$-$2\times10^{5}$ such
transients per year over the whole sky, or 10-200 above a threshold of
$10^{-8}$~\ecs. Connors et al. suggested that most events are stellar
flares from dMe-dKe stars, caused similarly as solar flares, namely by
energetic magnetic field reconfigurations.

\cite{ambruster1986} reports ten transients with the HEAO 1 A-1
experiment, eight of which are stellar flares and two are speculated
to be X-ray counterparts to faint GRBs. They infer an all-sky rate of
1500-3000 yr$^{-1}$ above $8\times10^{-11}$~\ecs\ (0.5-20 keV).

\cite{pye1983} reports twenty-seven hours-long transients with Ariel-V
SSI, eleven of which are identified with stellar flares and one is a
GRB. The remaining transients are speculated to be a mix of these
types of sources. The all-sky rate is inferred to be $\sim$1500
yr$^{-1}$ above 4$\times10^{-10}$~\ecs\ (2-18 keV).

\cite{gotthelf1996} searches through all IPC (Imaging Proportional
Counter) data in the 0.5-3.5 keV bandpass from the Einstein
Observatory for transients lasting shorter than 10 s and finds 42
cases with peak fluxes between 2$\times10^{-10}$ and
$1\times10^{-9}$~\ecs. This implies an all-sky rate of
$2\times10^6$~yr$^{-1}$ or 16 yr$^{-1}$ above
$1\times10^{-8}$~\ecs\ which appears comparable with the Connors et
al. result although there is a narrower selection for duration.

\cite{arefiev2003} combines data sets, investigating a 'few hundred'
FXTs from six instruments on five observatories that operated in the
1970s through 1990s with widely different sensitivities and fields of
view.  They conclude that FXTs comprise a variety of types of sources,
predominantly GRBs and stellar flares and that the origin of many,
particularly apparent GRBs without gamma-ray signals (which later were
called XRFs), are of unknown origin. Arefiev et al. estimate an
all-sky rate of tens of thousands per year above a fluence of
$10^{-2}$ Crab-sec.

Searches of FXTs at much fainter fluxes than mentioned above were also
carried out with X-ray narrow field telescopes aboard XMM, Chandra,
ROSAT, XRISM \citep[e.g.,][]{sun1998,
  quirola2022,tsuboi2024,eappachen2024,khan2025}. Several faint FXTs
with duration from about 100 s to 1000 s were found down to flux of
around 10$^{-13}$~\ecs. In this regime, a wider variety of sources are
making up FXTs. In addition to stellar flares and GRBs/XRFs dominating
the brighter end, faint FXT include tidal disruption events
\citep[e.g.,][]{jonker2013,auchettl2017}, supernova shock break out
\citep[e.g.,][]{waxman2007,alp2020,novara2020}, GRBs orphan afterglows
\citep[e.g.,][]{bauer2017,wichern2024} and quasi-periodic eruptions
from massive black holes
\citep[e.g.,][]{miniuti2019,giustini2020,quintin2023,wenbin2023}.

Currently, there are two work horses for detecting FXTs with wide
field monitors: MAXI (Monitoring of All-sky X-ray Image) on the
International Space Station \citep{matsuoka2009} and WXT (Wide-field
X-ray Telescope) on the Einstein Probe
\citep[EP,][]{yuan2022,yuan2025}. ISS and EP both lack gamma-ray
instruments. Therefore, MAXI and WXT lack synchronous gamma-ray
coverage and depend on such coverage asynchronously from elsewhere.
MAXI is particularly efficient in detecting stellar flares since these
generally last longer than the sample time of any point on the sky of
about 1 min per 1.5 hr. MAXI has for the past 15 years (up to October
10, 2025) reported through its alert messaging system 150 stellar
flares from 29 sources \citep[e.g.,][]{tsuboi2020}. WXT is perfectly
suited for detecting FXTs, thanks to its large instantaneous field of
view of 3600 square degrees and its deep sensitivity of
2.6$\times10^{-11}$~\ecs\ (0.8 mCrab) in a 1000 s observation in soft
X-rays (0.5-4 keV).  A preliminary analysis of the first year of EP
observations \citep{wu2025} resulted in 128 transients, that included
28 stellar flares, 30 events with a gamma-ray counterpart and 52
events without, either because they are XRFs or because no gamma-ray
instrument covered the event.  Another study \citep{zhang2025}
searched the gamma-ray counterpart in 63 FXTs detected by EP (that
excluded stellar flares and known sources), finding 14 such
events. While the identification of the full EP sample is ongoing,
extensive follow-up observations of a subset shows that several of
them are associated to GRBs, including high redshift GRBs, XRFs
\citep[e.g.,][]{liu2025,gianfagna2025,sun2025} but also some to TDEs
\citep{li2025} and oddballs \citep{zhang2025b}.

In the 1980s, FXTs were ill understood and in strong need of
exploration. Thanks to the prospering development of wide-field coded
aperture cameras with large-area multiwire proportional counters
\citep[e.g.,][]{proctor1978,fenimore1978,willmore1984,brinkman1985,skinner1987,mels1988},
SRON started in the mid-1980s the successful design and construction
of the Wide Field Camera (WFC) instrument for the Italian-Dutch
BeppoSAX X-ray observatory with the specific goal to collect data on
FXTs. The WFC was launched in April 1996. The mission lasted 6
years. The WFC was very successful in detecting thousands of transient
events and, for instance, made an essential contribution to solving
the GRB distance scale \citep{costa1997,jvp1997}. Much of the work
with the WFC has been published, but a specific paper reporting the
complete results of the FXTs detected by the WFC and their proposed
identification has been missing. The present paper aims to fill this
void, providing a complete flux-limited sample of FXTs that can be
used as a reference for current wide field monitor missions.

We have performed a systematic search of FXTs in the WFC data archive,
which, thanks to enormously improved computational capabilities in the
two decades since the operational end, has become much better
feasible. In addition, we provide simultaneous data from the Gamma-Ray
Burst Monitor (GRBM) on BeppoSAX, enabling for all these WFC FXTs a
homogenous coverage of their gamma-ray emission. We further supplement
their gamma-ray properties with data from the Burst And Transient
Source Experiment (BATSE) on the Compton Gamma-Ray Observatory when
available. We present some analyses of the catalog and compare them
with studies from other satellites. In Sect.~\ref{instr}, we describe
the employed instrumentation, in Sect.~\ref{obs} the observations and
in Sect.~\ref{search} the method of search for FXTs in WFC data. The
result is a catalog that is described in
Sect.~\ref{cata}. Section~\ref{ana} discusses the analysis of the
overall spectra and classification of the uncertain cases in the
catalog. Section~\ref{dis} presents a discussion of the results.  An
extensive analysis of the comparative properties of GRB, XRR and XRF
classes is forthcoming (Piro et al., in preparation).

\section{Instrumental description}
\label{instr}

BeppoSAX \citep{boella1997} hosted two identical cameras comprising
the WFC \citep{jager1997}, pointed in opposite directions and
perpendicular to the pointing of the three narrow field instruments
(NFI) onboard (LECS, MECS and PDS). They were coded aperture cameras,
consisting of a multi-wire position sensitive large-area
(25.6$\times$25.6 cm$^2$) proportional counter \citep{mels1988} paired
with, at a distance of 70 cm, a similarly sized gold-plated
stainless-steel mask etched with 20 thousand holes following a pattern
based on a triadic difference set \citep{zand1994}. This resulted in
the following characteristics: a bandpass of 2 to 30 keV, a photon
energy resolution of 20\% at 6 keV (full width at half maximum), a net
effective area of 141 cm$^2$ at 6 keV, a highest possible photon
detection rate of 2000 cts\,s$^{-1}$, a field of view 40×40 square
degrees (3.7\% sky coverage per camera), an angular resolution of 5
arcmin and a source location accuracy down to 0.7 arcmin (68\%
confidence). As a reference, the standard Crab X-ray source, when
placed on axis, had a photon-count rate of 280 cts\,s$^{-1}$ in the
full bandpass, while the count rate due to the cosmic diffuse X-ray
background was approximately 160 cts\,s$^{-1}$.

The coded aperture imaging yields coded detector images that need to
be decoded on the ground to obtain sky images. This imaging has the
advantage, compared to focusing techniques, that it enables a large
field of view and relatively cheap optics, but the disadvantage is
that photons of all celestial sources are detected on the same
detector area resulting in cross talk between different positions in
the field of view. The WFC sensitivity ranged between 15 mCrab (2-30
keV, integration time 1000 s) in the crowded Galactic center field
with many bright sources, particularly Sco X-1, and 6 mCrab far away
from the Galactic plane (5-sigma detection threshold). An in-depth
description of the WFC can be found in \cite{jager1997}.

The WFC instrument was very well suited for detecting FXTs because it
combined roughly 1-day exposures with a large field of view at
sufficiently high angular resolution to enable identification with
known X-ray sources or counterparts in the optical, radio or other
wavelengths.

The second instrument on board BeppoSAX relevant for the present study
is the GRBM \citep{amati1997,feroci1997,costa1998}.  It consisted of
four independent CsI(Na) detectors also used as lateral
anticoincidence shields of the Phoswich Detector System
\citep{frontera1997}, each with a size of 1136 cm$^2$, had a bandpass
of 40 to 700 keV and a sensitivity of $10^{-8}$~\ecs\ for 1 s. Two of
the four detectors were perpendicular to the viewing direction of a
WFC unit. Therefore, an event observed with a WFC camera had an
optimum GRBM collecting area.

Finally, we use publicly available data\footnote{BATSE data is
available at URL gammaray.nsstc.nasa.gov/batse/grb/} from the BATSE
instrument \citep{fishman1994,paciesas1999} on the Comption Gamma-Ray
Observarory (CGRO). This was an array of NaI detectors for the study
of GRBs that flew from 1991 to 2000 and had full sky coverage in the
nominal 20-325 keV bandpass.

\begin{figure*}[!t]
\centering
\includegraphics[width=1.5\columnwidth]{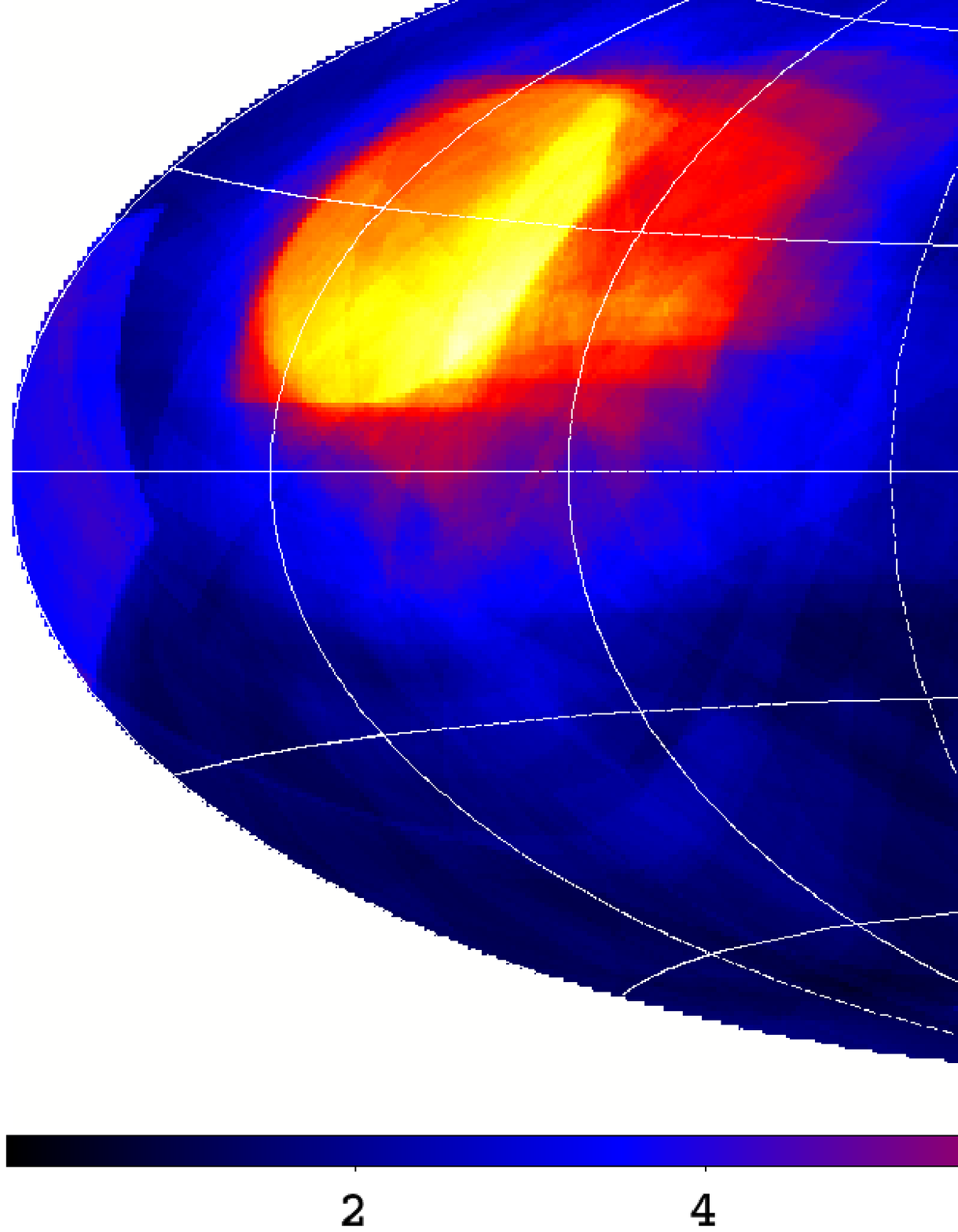}
\caption{Aitoff projection Galactic map of WFC exposure in units of Msec}
\label{fig1}
\end{figure*}
\begin{figure*}[!h]
\centering
\includegraphics[trim={0 1cm 0 2cm},width=1.8\columnwidth]{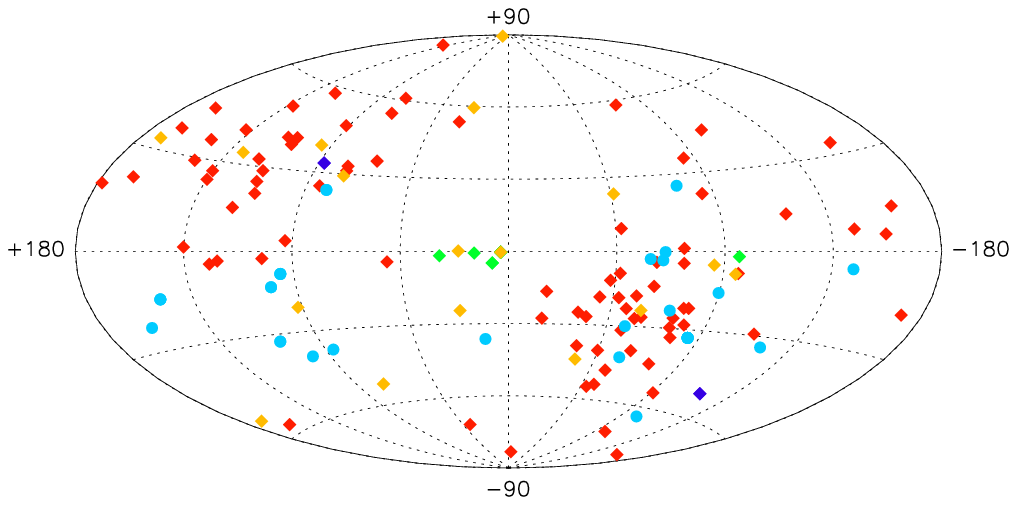}
\caption{Aitoff projection Galactic map of 149 FXTs. Red diamonds
  refer to GRBs, light blue to stellar flares, orange to unconfirmed
  GRBs ('GRB?' in Table~\ref{table1}), green-crossed o flares from X-ray
  binaries and dark blue to unconfirmed stellar flares}
\label{fig2}
\end{figure*}

\section{Observations}
\label{obs}

The BeppoSAX satellite was launched on April 30, 1996, in a low-Earth
low-inclination orbit with a height of about 600 km and an inclination
angle of 4 deg. The near to equatorial orbit ensured the least
disturbance by high-energy particle backgrounds from Cosmic Rays by
taking full advantage of the screening effect of the Earth's magnetic
field and avoiding most of the extreme particle flux in the polar
regions and the South Atlantic Anomaly (SAA) ), as demonstrated by the
Particle Monitor on board the satellite \citep{campana2014}. That
minimized the necessity to switch off the instruments during passages
of such high flux portions of the orbit \citep{boella1997}.  Regular
operations started in June and continued until April 30, 2002, for a
total of 2156 d or 31 thousand satellite orbits. Apart from
intermittent short periods due to operational incidents, the science
instruments were turned off for longer periods during May-August 1997
to install the one-gyro mode, and during October 2001 to install the
gyro-less mode. Furthermore, WFC unit 1 was turned off in May 1998 to
let it recover from presumable frost bite. The total net time that the
high voltage (HV) of the proportional counters was at a normal
operational level was 1285 days for unit 1 and 1353 days for unit
2. These numbers are impacted by the fact that not always sufficiently
accurate star tracker data is available to infer the pointing
information of the WFC within a few arcminutes or ensure pointing
stability. This reduces the useful observation time by 35\%. Finally,
the nearby Earth often obscures (parts of) the field of view. Only
44\% of the time the Earth is completely outside the field of view of
a WFC camera.

Both WFC cameras continuously observed the X-ray sky, except for times
when the Earth blocked the field of view or when the satellite passed
the SAA. In the latter case, the cameras were put in standby mode by
lowering the high voltage. Including times when the Earth occulted
(part of) the field of view, the sum of the exposure times of both
cameras when the pointing was stable and known is 5.33 years or
approximately 168 million seconds. In Fig.~\ref{fig1}, a Galactic map
of the coverage is given.  The WFC did not provide a uniform coverage
of the sky. In particular, the maximum coverage was obtained on the
ecliptic poles (due to absence of Earth occultations there) and is
approximately 13 million seconds. Also, fields around the Galactic
center, the Galactic plane, the Crab nebula (opposite the Galactic
center), Cygnus and LMC are more exposed. The Galactic center and Crab
nebula fields were the sole primary targets for the WFC among the
BeppoSAX instruments.

The WFC sensitivity ranged from 10$^{-9}$~\ecs\ (2-10 keV) for a time
scale of 10 s, to a few times 10$^{-11}$~\ecs\ for a time scale of
10$^5$~s. This implies that it was not sensitive to detect flashes
like those discovered by \cite{gotthelf1996} in Einstein data or
\cite{sun1998} in ROSAT data. On the other hand, the WFC sensitivity
is comparable to that of EP WXT for events shorter than about 50-100
s, while it does not reach the EP-WXT sensitivity for longer durations
(see Sect.~\ref{dis}).

The GRBM was particularly used to detect GRBs. When a threshold was
exceeded, the instruments triggered into high-resolution data
collecting mode. The GRBM catalog of GRBs includes 1082 GRBs, one
third of which was missed by the on-board trigger logic for various
reasons and identified on ground using more sensitive algorithms
\citep{frontera2009}. Apart from GRB data, the instrument provided
rate measurements at 1 s resolution and two bands: nominally 40-700
and $>$100 keV. For most of the FXTs reported here, these rate
measurements are available. Of the 100 putative GRBs detected by the
WFC, 79 were detected by the GRBM.

The Large Area Detectors (LADs) of the BATSE instrument were slightly
more sensitive than the GRBM, particularly towards lower energies (the
bandpass went down to 20 keV). They operated simultaneously with
BeppoSAX for 3.8 yr, ending with the termination of the Compton Gamma
Ray Observatory on 4 June 2000. Many of the events which were visible
to BATSE did not activate its onboard trigger system which indicates
that the 50-300 keV peak fluxes were below 0.2 cts\,s$^{-1}$cm$^{-2}$
(1024 ms time scale). Like GRBM, BATSE was only sensitive to
burst-like features with strong variability below 1 min time
scales. Of the 65 GRBs detected by WFC before the end of CGRO, 27 GRBs
were detected by BATSE through time-wise coincidence. The remaining
were probably mostly occulted by Earth as seen from CGRO in its
low-Earth orbit.

\section{Search method and selection criteria}
\label{search}

For our purposes, a transient is a point source in the sky which, for
a limited time ($\sim$1 s -- $\sim$1 d in our case), brightens to a
flux level significantly higher than the noise at that sky location as
caused by all other sources (other astrophysical point sources in and
outside the field of view, the cosmic diffuse background and the
particle-induced background). From the complete WFC database, we
extracted sky images in the complete bandpass for data stretches of 10
s, 60 s, 300 s and 1 BeppoSAX orbit (1.5 hr) duration, sampling the
data at intervals half the stretch durations. The sky images were
reconstructed from coded aperture images through normalized cross
correlation of the detector images with the mask pattern, combined
with Iterative Removal Of Sources
\citep[IROS;][]{hammersley1986,jager1997}.  The detector spatial data
were binned to match the coded mask resolution of 1$\times$1
mm$^2$. Since the mask consists of 255$\times$257 bins, the cross
correlation results in images of 510$\times$512 pixels. The sky images
were normalized to the standard deviation from the expected Poisson
noise, so that they are in units of significance.  These images were
searched for FXTs

\begin{enumerate}
\def\labelenumii{\alph{enumii}.}
\item
    on catalog positions of, for instance, flare stars within 100 pc
    in the lists of \cite{kowalski2024} and of stars that have shown
    flares in MAXI observations (29 stars). The trigger criterion for
    a candidate transient then is a significance of 3 or higher;
\item
    through blind searches in the whole field of view. The trigger
    criterion for a candidate transient then is a significance of 5 or
    higher.
\end{enumerate}

In all practical circumstances here, significance is equal to the
signal-to-noise ratio because the source contribution to the Poisson
noise is much lower than from all other sources. A significance of 5
or higher yields a probability of less than 7.5\% that there is one
pixel in a single image with such a significance from noise.

The validation of candidate transients to true transients involved the
visual check of the sky image for errors in the reconstruction (e.g.,
high noise levels for parts of the image) and, subsequently,
generating light curves for the detected positions at time resolutions
of 1, 10, 60, 300 and 900 s. If the light curves showed a clear
temporary enhancement of the flux 3$\sigma$ above the zero level over
multiple time bins within the original 5 or 3$\sigma$ triggering time
interval, the trigger was validated as a transient. For new sources,
this reduces the false source probability to less than
2$\times10^{-5}$ (chance probability for statistical $>$3$\sigma$
positive fluctuations in two or more light curve bins for that
particular position in the image, times roughly 10 trials in the light
curve).

There are three caveats to this method. First, the calibration of the
detector spatial response is not complete. On local scales, there are
deviations from a flat response.  This is typical for multi-wire
proportional counters. It yields an overall systematic localization
error of 0.7 arcmin (68\% confidence).  This imperfection makes it
impossible to predict the image position of any point source any
better than that, even if the celestial position of sources in the
field of view is known at sub-arcsec accuracy like for most catalog
sources. Thus, when searching for detections, the point-spread
function is fitted to any high signal-to-noise pixel with X and Y left
free within boundaries of +/- 3 pixels in our case.  Obviously, this
will bias the fitted flux to positive and negative stochastic noise
fluctuations. Therefore, it makes sense to increase the significance
thresholds by a factor of $\sqrt{3}$, so to 5.2 for catalog sources.

A second caveat is that employing data stretches of 10 s, 60 s, 300 s
and one BeppoSAX orbit (often with an exposure of about 1800 s) leaves
less than optimum detection thresholds for transients with durations
in between these values. We estimate that the maximum significance
disadvantage of this sampling is about 30\%.

A third caveat is that the point-spread function (PSF) full-width at
half maximum (FWHM) is between 1.0 sky pixel at best and a few pixels
for far off-axis angles and large spectral hardness. Thus, for soft
spectra and small off-axis angles, the PSF may be somewhat
undersampled which deteriorates the sensitivity.

Apart from the searches for FXTs in images, searches were also
performed in the time profiles of detector event rates at 1 s
resolution, for whole detectors as well as for quarters of
detectors. This was done automatically as well as by eye.  This search
method is sensitive for burst-like (sub-minute) events; for longer
time scales the event rates are subject to changes in particle
background and Earth obscuration. When a candidate event was found,
its reality was checked by generating a sky image for the duration of
the event and searching for an accompanying point source like
above. False events are usually due to solar activity (in extremely
bright X-rays or particle winds) outside the field of view or to X-ray
bursts from active low-mass X-ray binaries
\citep[i.e.,][]{galloway2020}.

The procedure for FXT identification evolved over years of working
with WFC data. It has a strong human component to it (i.e., the many
visual checks). Roughly 90\% of all candidates were rejected during
the validation. It may be that some faint transients are not caught
but we note that the detection threshold obtained is close to the one
predicted by \cite{jager1997}, see Fig.~\ref{fig7}.

\section{Catalog}
\label{cata}

Our search resulted in 149 FXT detections. Table~\ref{table1} presents
their basic parameters. The column description is as follows.

\begin{enumerate}
\def\labelenumi{\arabic{enumi}.}
\item
  ` ID' - Identification, following the GRB convention of date at which
  the transient event occurred supplemented, for multiple events on a
  single day, with a rank letter from the alphabet;
\item
  `Ref' - `u' for unpublished event, `a' for published by \cite{dalessio2006}.
  Numbers refer to the papers cited in the footnote;
\item
  'r' - an x denotes a case in which the community was alerted in near
  real time (i.e., within hours) with the position of the event (55
  cases). A + sign denotes a case where additionally the NFI observed
  the event within hours (38 cases\footnote{See URL {\tt
    www.mpe.mpg.de/$\sim$jcg/grbgen.html} and references therein});

\item
  `OP' - BeppoSAX observation identification (`Observation Period');
\item
   WFC unit that detected the event (1 or 2);
\item
  MJD(UTC) of the start of the event;
\item
  T90 duration. This type of duration was introduced in
  \cite{fishman1994}. We determined it by fitting a cubic spline
  through the light curve and determining for the cubic spline, as a
  function of time from the start time of the event onward, the
  fraction of the total cts\,cm$^{-2}$ fluence accumulated. T90 was
  taken as the duration between the times when 5\% and 95\% of the
  fluence is accumulated;
\item 
  1$\sigma$ error in T90. This was determined by creating 1000 Monte
  Carlo realizations of the light curve, sampling from a normal
  distribution for each data point a flux from the measured value and
  its 1$\sigma$ uncertainty, and calculating the standard deviation of
  the 1000 T90 values;
\item
  R.A. (eq. 2000.0) of the event, as determined from the centroid of the
  fitted point spread function and the attitude data of BeppoSAX. The
  position has been determined from the full-bandpass image with the
  optimum signal-to-noise ratio;
\item
  Dec. (eq. 2000.0), determined as for R.A.;
\item
  The positional uncertainty of the event, expressed as the radius of
  the 68\%-confidence circular error region, in units of
  arcmin. This is according to the calibration as discussed in
  \cite{heise1998};
\item
  The angular separation between the event position (columns 9 \& 10)
  and the identified optical counterpart (last column) in arcmin. If
  there is no identified counterpart, as is the case in a number of
  GRBs without optical follow-up, this column is kept empty;
\item
  CGRO-BATSE counterpart id. \citep[for definiton,
    see][]{fishman1994,paciesas1999}, or offline identification
  (`untrig.').  Note that BATSE ceased operations on June 4, 2000;
\item
  WFC (2-30 keV) peak fluxes in cts\,s$^{-1}$cm$^{-2}$ and dead time
  corrected. Note that the WFC flux for the Crab source is 2.1
  cts\,s$^{-1}$cm$^{-2}$, which for a Crab-like spectrum translates to
  2.0 and 3.3$\times10^{-8}$~\ecs\ for 2-10 and 2-30 keV,
  respectively;
\item
    GRBM (40-700 keV) peak fluxes in cts\,s$^{-1}$ per detector
    corrected by division with the cosine of the off-axis angle. Note
    that 1) in case the GRBM peak flux is left blank, it was not
    possible to reliably infer a value because the transient lasts
    $\geq1$~h which makes it impossible to distinguish it from changes
    in the background that have a similar time scale; 2) in case the
    GRBM peak flux is marked by 'n.a.', GRBM data is not available; 3)
    this is a WFC-based table, implying that in a number of cases the
    GRBM did not detect the event. The GRBM peak rate is not corrected
    for dead time, which introduces an error of 0.3\% at maximum;
\item
  The spectral softness ratio between the WFC (2-30 keV) and GRBM
  (40-700 keV) peak fluxes, multiplied by a factor of 1000. Note that
  the peak fluxes in the two instruments are measured at the same time
  resolution as employed in Fig.~\ref{figlc2} but not necessarily at
  the same time;
\item
  Type of event, based on the optical/radio counterpart (next column)
  or when there is a clear simultaneous gamma-ray signal. When there
  is no clear counterpart and no clear gamma-ray signal in either GRBM
  or BATSE, the type is labeled with a question mark.
\item
  Identification of the event with the most likely counterpart. This
  column may also have facts of interest. One of these facts is
  whether \cite{dalessio2006} finds whether a GRB is X-ray rich
  (`XRR') or XRF. Note that we have a different criterion for
  XRF (see Sect.~\ref{dis}).
\end{enumerate}

A synopsis of Table~\ref{table1} is the following. 149 transient
events are listed of which 64 have not been published before.  100
events are associated with GRBs \citep[of which 23 unpublished;
  c.f.][]{vetere2007}. 70 are without GRBM counterpart (49 stellar
flares and 21 others). 11 stellar flares are associated with
supergiant fast X-ray transients (SFXTs) or other high-mass X-ray
binaries (HMXBs) and 38 from flaring stars. Nine sources (2 SFXTs, 7
stars) are repeating between 1 and 6 times. The total number of unique
sources in the table is 127, out of which 22 are stars (including 5 RS
CVns, 3 BY Dra variables, 3 T Tau stars and 2 G-type stars), 4 SFXTs,
1 HMXB and 100 GRBs/XRFs. 18 events have tentatively been identified
as GRBs, but without the detection of a gamma-ray signal by the GRBM,
BATSE or any other instrument, thus falling under the XRF class.  Two
of these (020410 and 020427) have detected afterglows which makes
their association to GRBs more probable than the remaining 16
cases. Those 16 cases have been labeled 'FXT' in the last column of
Tables~\ref{table1} and \ref{table2}, and in Fig.~\ref{figlc2}.

Figure~\ref{fig2} shows a Galactic map of all 149 transients. The
distribution reflects the non-isotropic coverage of WFC observations,
see Figure 1.

Figure~\ref{figlc1} shows the light curves of 49 transient events that
can be attributed to stars and X-ray binaries. All data are shown of
subsequent observations that continuously pointed at the
source. Usually this concerns one OP, but occasionally this concerns
up to 5 OPs. For all data we applied the standard allowed star tracker
configurations. Furthermore, we included in this data set those data
where the source radiation traverses at least 150 km above the Earth
horizon. During some of these data, the Earth may obscure other parts
of the field of view. For the light curve of event 010315 (GT Mus), a
further constraint was applied where the Earth was required to be
completely out of the field of view, as in this particular case the
standard star tracker configuration was unacceptable when the Earth
was in the field view (the flux of GT Mus would drop to zero during
these times).

Figure~\ref{figlc2} shows the light curves of the 100 GRBs, focusing
on times closer to the (shorter) events. Also shown are the 5-30/2-5
keV spectral hardness ratio and the GRBM photon count rates. The
latter data are raw data without background subtraction, in contrast
to the WFC light curves. For further info on GRBM detected events
(e.g., fluence and peak flux in cgs units), see \cite{frontera2009}
and \cite{guidorzi2011}. The same selection of good time intervals was
applied as for Fig.~\ref{figlc1} (allowed star tracker configurations
an a source that is at least 150 km above the Earth horizon).

We note that very bright events (i.e., 10 Crabs or brighter) were
excluded on 971010 (MJD 50731.505104, OP2622, WFC unit 2), 000305
(51608.175961, 8616, 1) and 000411 (51645.119576, 8902, 2) because
they exhibit a suspect top-hat light curve with a duration of 10-16
s. These are suspected to be false because of improper processing of
raw satellite data.

\setcounter{figure}{4}
\begin{figure}[b]
\includegraphics[height=\columnwidth,angle=270]{aa58641-25-f5a.ps}
\includegraphics[height=\columnwidth,angle=270]{aa58641-25-f5b.ps}
\caption{Spectra showing strong Fe-K emission lines for CC Eri and SAX
  J1819.3-2525 on top of power-law continuum models. The top panels
  show the data and model spectrum, the middle panels the residuals
  between data and model and the bottom panels the same residuals
  after setting the line fluxes to zero. The equivalent widths of
  these two cases are 2.17 and 1.80 keV, respectively. The line widths
  have been fixed at 0 keV.  The line of SAX J1819.3-2525 has been
  discussed in \cite{zand2001}.}
\label{fig5}
\end{figure}

\section{Analysis}
\label{ana}

\setcounter{table}{2}
\begin{table}
\centering
\caption{Five most significant iron line detections in stellar flares.}
\label{table3}
\begin{tabular}{lllll}
Event & Source             & E(line) & EW  & Signi- \\
      &                    &(keV)    &(keV)& cance \\
\hline
970123 & BY Dra            & 6.52$\pm$0.13 & 1.63 &  5.2 \\
970321 & NLT 51688         & 6.53$\pm$0.07 & 1.89 &  4.7 \\
990909 & SAX J1819.3-2525  & 6.45$\pm$0.05 & 1.80 & 12.7 \\
001207 & CC Eri            & 6.53$\pm$0.05 & 2.17 & 15.0 \\
010119 & Algol             & 6.68$\pm$0.10 & 2.10 &  5.9 \\
\hline
\end{tabular}
\end{table}

We extracted overall WFC spectra of all 149 events, integrating over
the duration of each event. These spectra were fitted with 3 models: a
power law, a black body spectrum and thermal bremsstrahlung.  All
spectra were absorbed with XSPEC model `phabs', with cross sections
according to \cite{verner1996}, abundances according to
\cite{wilms2000} and equivalent Galactic hydrogen column densities
$N_{\rm H}$ following \cite{hi4pi2016} through {\tt ftool} 'nh'
(version 6.36). A narrow emission line was included at the energy of
the Fe K line. For most spectra, the fits with the power law are
best. 11\% of the black body fits show lower $\chi^2_{\rm r}$ and 39\%
of the bremsstrahlung fits. However, in all these cases the quality of
the fits is similar for the power law fits. Therefore, for the
remainder we only consider the power law fits. Interstellar absorption
and a narrow line were set to Galactic values and zero, respectively,
except for a few cases where they were necessarily left free to obtain
a good fit. Table~\ref{table2} presents the results of the spectral
analysis with a power law, including a presentation of the 2-30 keV
fluence and 2-30 keV isotropic energy output for all cases for which a
distance is known.  Three groups are discernable in distance and,
therefore, energy output: nearby stellar flares ($10^{34}$-$10^{38}$
erg), flares from X-ray binaries at a few kpc ($\sim10^{40}$ erg) and
GRBs ($\sim10^{51}$ erg). See Tables~\ref{table3} and \ref{table4} for
cases where an emission line was included at the Fe K location and
$N_{\rm H}$ was left free, respectively.

For 5 stellar flares the inclusion of the emission line at
$\approx$6.5 keV was necessary for a good fit. The 5 cases are in
decreasing order of significance: 001207 (flare from CC Eri), 990909
(first flare from SAX J1819.3-2525), 010119 (third flare from Algol),
970321 (flare from 1RXS J214641.1-854303) and 970123 (first flare from
BY Dra), see Table~\ref{table3}. The two most significant Fe-K lines
are shown in Fig.~\ref{fig5}. Note that none of the GRB average
spectra showed a significant Fe-K emission line.  However, some bursts
have transient narrow spectral features in WFC data: GRB990705
\citep{amati2000}, GRB990712 \citep{frontera2001} and GRB011211
\citep{frontera2004}.

For 5 FXTs, it was necessary to leave $N_{\rm H}$ free during the
spectral fitting, see Table~\ref{table4}. These all happen to be cases
where the event is from a HMXB. Four of these are from SFXTs. This is
well understood. In these systems, variable winds can temporarily
obscure the line of sight, giving rise to changes in $N_{\rm
  H}$. There is one GRB (GRB000528) that shows evidence of a changing
high hydrogen equivalent column density \citep{frontera2004b}.

\begin{table}
\centering
\caption{Excess values for $N_{\rm H}$ if left free.}
\label{table4}
\begin{tabular}{llcc}
Event & Source& Excess $N_{\rm H}$ & Galactic $N_{\rm H}$ \\
      &       & \multicolumn{1}{c}{(10$^{22}$~cm$^{-2}$)} \\
\hline \\
960825  & IGR J17544-2619 1  & 22.0$\pm$1.2 & 1.32\\
980311  & SAX J1818.6-1703   & 161$\pm$85   & 1.17 \\
990220a & SAXJ 1819.3-2525 3 & 23$\pm$15    & 0.03 \\
000920  & IGR J17544-2619 4  & 3.2$\pm$2.7  & 1.17 \\
010324b & AX J1845.0-0433    & 26$\pm$11    & 1.40 \\
\hline
\end{tabular}
\end{table}

\begin{figure}
\includegraphics[trim=0.cm 0cm 0 0,clip,width=1.05\columnwidth]{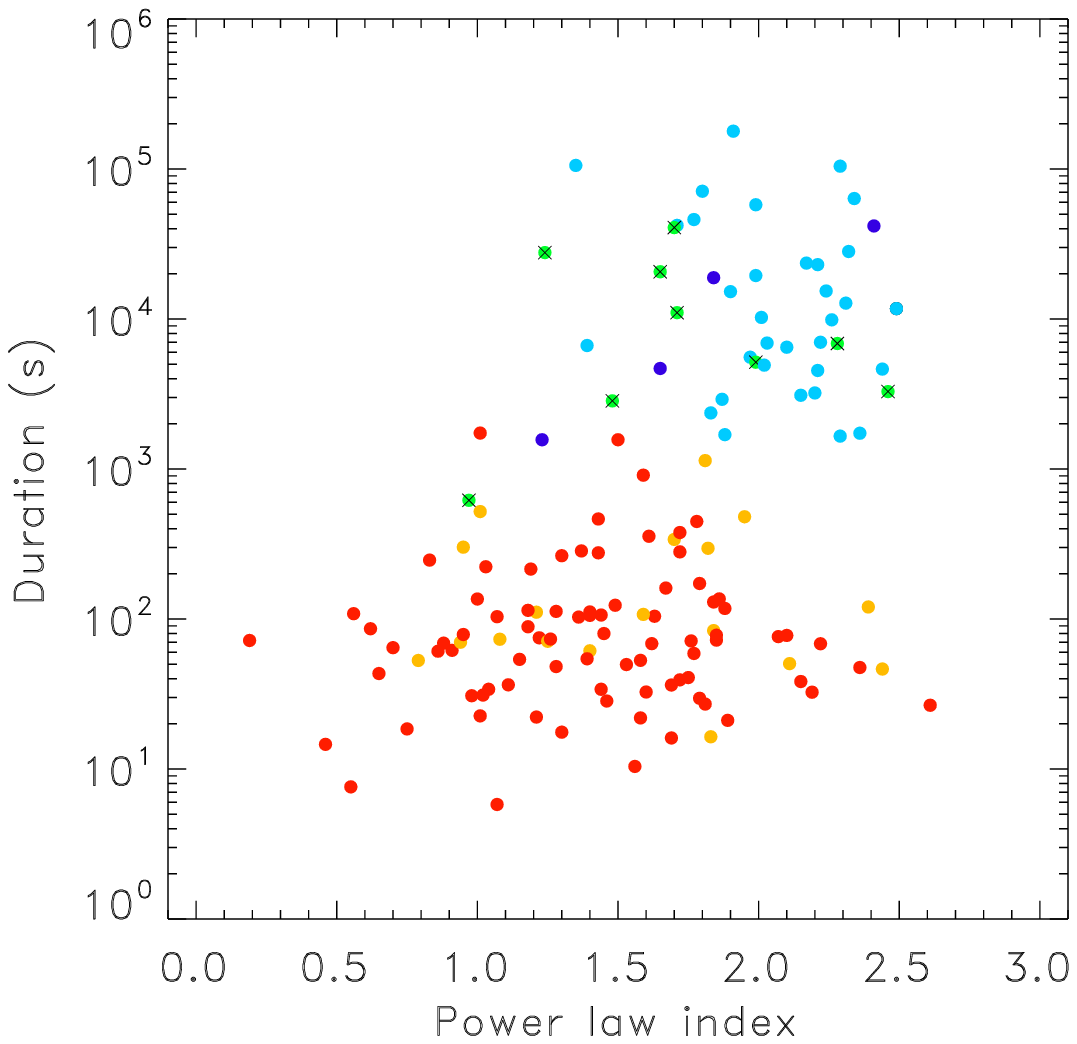}
\caption{Diagram of T90 duration versus power law photon index for all
  WFC events. Red dots refer to GRBs, light blue to stellar flares,
  orange to unconfirmed GRBs ('GRB?' in Table 1), green-crossed to
  flares from X-ray binaries and dark blue to unconfirmed stellar
  flares.}
\label{fig6}
\end{figure}

Figure~\ref{fig6} shows the diagram of T90 duration (column 7 of
Table~\ref{table1}) versus power-law photon index (column 7 of
Table~\ref{table2}). The 5 different types of sources are indicated
with different colors. GRBs and stellar flares are clearly distinct in
this diagram. GRBs are short and hard (low photon index), while
stellar flares are the opposite. The uncertain classes are mostly
associated with GRBs without gamma-ray signals (`GRB?' in
Table~\ref{table1}, orange symbols in Fig.~\ref{fig6}).

Since the uncertain GRB cases are all without a GRBM signal or data,
they did not trigger real-time alerts to the community and follow-up
observations with, among instruments elsewhere, the NFI. Thus, the WFC
data are the only data on these events. There is one exception,
GRB991217, which triggered a prompt reaction because of the strong WFC
signal and a noise peak in the GRBM data. No NFI observation took
place but optical follow up was engaged with 1 m class telescopes. No
afterglow was detected brighter than $R\approx22$ within 0.5-2.5 d
\citep{mohan1999}.

Considering that the brightest X-ray stellar flares have a luminosity
of 10$^{32}$ erg~s$^{-1}$, this implies that stellar flares can be
detected by the WFC (taking a sensitivity of
2$\times$10$^{-10}$~\ecs\ for duration of about 1000 s) to a maximum
distance of 64 parsecs, which ensures the completeness of star
catalogue \citep{gaia2023}. The association of the uncertain GRBs to
X-ray-dominated GRBs, i.e. XRFs, is supported by the fact that these
events share the same duration-spectral index region populated by GRBs
(Fig.~\ref{fig6}).

There are a few unidentified longer cases that are on the border
between GRBs and stellar flares. They are the two events from GR Dra
and the two events from SAX J0346.6-3906 (dark blue symbols in
Fig.~\ref{fig6}). These are further discussed in Sect.~\ref{specials}.

\subsection{Interesting individual cases}
\label{specials}

\subsubsection{GR Dra}

\cite{williams1995,williams1997} reported the detection in November
1992 with CGRO-Comptel of the MeV transient GRO J1753+57. The source
was detected in two consecutive 14-d 'viewing periods' (VPs) with a
1-3 MeV peak flux in one VP of $10^{-6}$ cts\,s$^{-1}$cm$^{-2}$.  This
flux equals half the flux of the Crab source in the same bandpass.
The spectrum is consistent with that of an AGN. Within the 4 deg error
radius of the source (4-sigma confidence), 14 candidate counterparts
were identified which are not galaxy clusters. 3 of them are stars, 7
are AGN and 4 are of unknown type. By virtue of the similarity in
gamma-ray behavior to 2 other transients, the source was suspected to
be of AGN origin. \cite{carra1997} searched for optical counterparts
among 9 radio sources and 10 AGN in the error region but was
unsuccessful in finding an unambiguous variable counterpart.

We find two transient events in our catalog whose positions are
consistent with that of one of the candidates identified by Williams
as counterpart: 1ES 1727+590 = GR Dra. This object is, at 3.6 deg from
the centroid position of GRO 1753+57, the third closest of nine.
Therefore, we conclude that we have detected the second and third
outburst of the MeV transient GRO 1753+57 but now at 2-30 keV instead
of 1-10 MeV and that it is a Galactic star rather than of
extragalactic origin. The WFC-detected flares are much shorter than 4
weeks as observed by CGRO-Comptel. This suggests that the transient
event seen by CGRO may comprise multiple events in 2 VPs.

The peak flux that we observe, about 0.08 cts\,s$^{-1}$cm$^{-2}$, is
about 40 mCrab. Thus, the 0.5 Crab flux measured by CGRO suggests a
hard MeV tail. The X-ray spectrum is rather soft, with a photon index
of 2 or a bremsstrahlung temperature of 4.6 keV.

What type of star is GR Dra? It is a bright (V=8.26) G0 star at a
distance of 139 pc. \cite{schachter1996} detects optical variability
from this star. GR Dra exhibits persistent X-ray emission at a level
of 1.4 mCrab \citep[0.1-2.4 keV;][]{voges1999}. The astrometric excess
noise measured with GAIA is 0.134 mas \citep{gaia2023} which is too
low to detect binarity \citep[c.f.,][]{gandhi2022}.

If GR Dra is the same as GRO 1753+57, how can a G0 star give rise to MeV
transient emission? We suggest that GR Dra may house a compact object.
Sensitive follow-up observations should be interesting.

\subsubsection{SAX J0346.6-3906 = UCAC4 255-003783}

We see two transient events lasting less than 1 hr and peaking at
about 100 mCrab and with a moderately hard spectrum (photon index
1.8). There are no known X-ray sources at its location. There is one
optical counterpart: UCAC4 255-003783, a cool star in the Epsilon Cha
star association. Since SAX J0346.6-3906 is a repeating source, it
cannot be a (long) GRB. The event may be a faint X-ray burst due to a
distant X-ray binary. For a maximum expected luminosity, the Eddington
limit of a H-rich accreting neutron star, the distance would be
$\sim$30 kpc which would place it at a somewhat unlikely location
outside the Galactic Bulge.

\subsubsection{1RXS J214641.1-854303 = NLTT 51688}

This object has been identified with a V=13.4 M3.5 dwarf star NLTT
51688 \citep{riaz2006} at a distance of 16 pc \citep{gaia2023}, which
has never been seen to flare before but whose spectral type is fully
consistent with flaring activity. We detect a significant Fe-K line
from this source, which suggests that substantial obscuration of the
flare source occurs \citep[c.f.,][]{mushotzsky1993}.

\subsubsection{2RE J051724-352221 = CD-35 2213}

This object has been identified with a V=11.7 M4Ve dwarf star \citep{scholz2005}
at a distance of 12 pc \citep{gaia2023}, which has never been
seen to flare before but whose spectral type is fully consistent with
flaring activity.

\subsubsection{Others}

\textbf{971019}. This is the brightest and the softest GRB event. A
combined WFC/BATSE spectral fit over the whole event shows a peak energy
in the $\nu$F$\nu$ spectrum of 19±1 keV compared to an average peak energy of
about 200 keV for classical GRBs \citep{kippen2003}. This is the
prototypical X-ray flash (XRF) case of our sample.

\vspace{3mm}\noindent \textbf{980306} is classified as XRF on the
basis of BATSE data instead of GRBM data \citep{kippen2003}. It was
followed up with the BeppoSAX NFI on August 1-2, 1999, 17 months after
the event. This observation is not yet discussed in the literature. No
source was detected with MECS within the WFC error box above
1.7$\times10^{-3}$~cts\,s$^{-1}$ (3.5-sigma upper limit) corresponding
to a flux of $\sim10^{13}$~\ecs.

\vspace{3mm}\noindent \textbf{980429}. This is the first XRF that was
followed up with a sensitive device within a matter of days
\citep{heise1998b}. The NFI observed the position between 32.45 and
53.4 hrs after the flash and detected one source at a position of
R.A.$_{2000.0}$=130.1507 deg, Decl.$_{2000.0}$=+22.8584 deg with an
uncertainty of 1\farcm0. This is 1\farcm1 from the WFC centroid, so
fully consistent. The source is weak, with a MECS photon count rate of
(2.11$\pm$0.39)$\times10^{-3}$~cts\,s$^{-1}$ over 37364 s net exposure
time, and no variability nor spectrum could be detected.  The photon
rate implies a 2-10 keV flux of a few times $10^{-13}$~\ecs.  Since
there is no clear decay during the NFI observation, the source cannot
unambiguously be identified as the afterglow. Optical follow up after
2.8 d failed to find a counterpart, with limiting magnitudes of
B$\sim$22 and R$\sim$21 \citep{djorgovski1998}. This source was also
followed up in 2007 and not detected above an upper limit of
4.5$\times10^{-14}$~\ecs\ \citep[presumably in 0.5-10
  keV;][]{romano2008}.

\vspace{3mm}\noindent \textbf{991106}. This event cannot be completely
ruled out as a type-I X-ray burst \citep{cornelisse2002}. The Galactic
latitude of this event is -2.6 deg. The event was followed up with the
NFI between 7.9 and 27.0 hr after the event. The MECS exposure time is
31.7 ksec, and the LECS exposure 11.8 ksec. The MECS data show a
detection of a point source in the WFC error box. The flux is
1.8$\times10^{-3}$~cts\,s$^{-1}$ in the MECS, which for a Crab spectrum
is roughly 10$^{-13}$~\ecs\ which is very low for an X-ray burster
with such a short burst.

\begin{figure}
\includegraphics[trim=0cm 0cm 0 -2mm,clip,width=1.0\columnwidth]{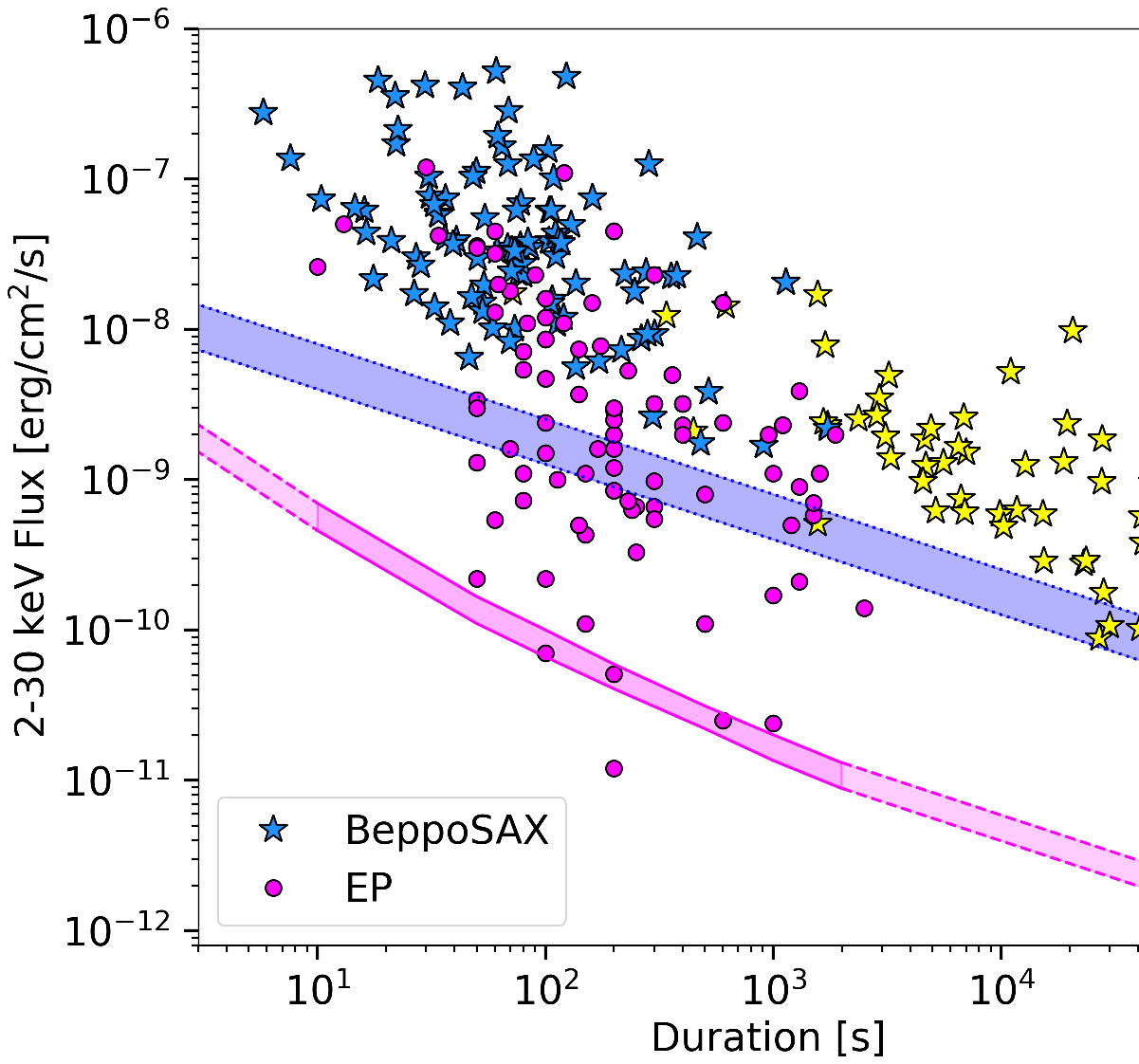}
\caption{Comparison of average 2-30 keV fluxes for Einstein Probe
    and BeppoSAX events. 149 BeppoSAX-WFC GRBs and stars from the
    present catalog are represented in blue and yellow asterisk symbols,
    respectively. 84 EP-WXT transients (without stars and Galactic
    sources, from \citet{wu2025} and \citet{aryan2025}) are in
    magenta-filled circles. The BeppoSAX-WFC sensitivity is drawn as a
    blue band since it depends on the particular X-ray source
    configuration in any observed sky image (for example, it depends
    on the inclination angle and on the pointing). The EP-WXT
    sensitivity is represented as a magenta band and corresponds to
    the 90\% confidence interval (as reported at
    {\tt https://ep.bao.ac.cn/leia/cms/article/view?id=38}) in the
    time range 10-1000 s. Below and above this time range, we extend
    the sensitivity lines assuming a photon limited regime ($t^{-1}$),
    and a background limited regime ($t^{-1/2}$), respectively (dashed
    lines). We note that the EP-WXT sensitivity was extrapolated to
    the 2-30 keV band assuming the same spectrum as here, so it is not
    to be considered as exact since the type of spectrum can change
    substantially the flux level.}
\label{fig7}
\end{figure}

\section{Discussion}
\label{dis}

The faintest FXT that we find has a peak WFC flux of
0.02~cts\,s$^{-1}$cm$^{-2}$ (and a duration of 20 h) which, for a
Crab-like spectrum, translates to $2\times10^{-10}$~\ecs\ (2-30
keV). For a 10$^3$ s exposure, the formal 5-sigma sensitivity is the
same value (2-30 keV) for fields away from the Galactic center and for
the on-axis position. For other positions in the field of view, the
sensitivity reduces approximately pyramidically towards the edge and
the average over the field of view is about $4\times10^{-10}$~\ecs.

Our coverage is 5.33 yr exposure for 2 WFC cameras times the 0.47 sr
field of view of one WFC camera. This equals 2.50 yr sr and implies a
yearly all-sky rate of 755$\pm$62 yr$^{-1}$. This is of the same order
of magnitude as what was found with the detections by Ariel V and HEAO
in the 1970s based on fewer events \citep{ambruster1986,pye1983}.

An interesting comparison to make is with the recent results from the
WXT on EP. From General Coordinate Network (GCN) circulars of the
first year of EP operations, \cite{wu2025} collected 128 FXTs,
including stellar flares and known sources, while \cite{aryan2025}
collected 72 FXTs, excluding stellar flares. For a field of view of
1.1 sr and assuming an observation efficiency of 90\%, this yields a
total all-sky rate of GRBs and stellar flares of roughly 1600$\pm$140
yr$^{-1}$ (note that the fraction of GRBs and stellar flares in this
sample is unknown at the writing of this paper). This is a factor of 2
higher than for the WFC. This can roughly be explained by a difference
in sensitivity. The WXT sensitivity is
0.26$\times10^{-10}$~\ecs\ (0.5-4 keV) for an exposure time of 10$^3$
s. EP-WXT has a substantially better sensitivity than BeppoSAX-WFC for
events lasting longer than about a few tens of seconds, because the
WXT is photon limited, it being a direct focusing camera, while the
WFC is background limited. In Fig.~\ref{fig7}, we plot the average
flux in the 2-30 keV range versus duration for events detected by
either WFC or WXT. Here, the EP-WXT flux has been extrapolated from
0.5-4 keV to 2-30 keV using the published WXT spectra and, therefore,
does not take into account any possible spectral breaks between 4 and
30 keV. For the duration, we applied the T90 values for the WFC sample
and the durations mentioned in the appropriate GCN circulars for the
EP sample. When we consider only the 50 EP events above the
BeppoSAX-WFC sensitivity (taking the center line of blue-filled region
as dividing line), the fraction of EP events that would be detectable
by the WFC is 0.61. If we correct the value of 1600 above for this
fraction, the all-sky rates of transients are comparable: the EP rate
reduces to 960$\pm$221, a factor of 1.27$\pm0.24$ higher than the WFC
rate.  Systematic effects, for instance due to the different
bandpasses of the WFC and WXT, will add uncertainty to this number.

The comparison of BeppoSAX versus EP events in Fig.~\ref{fig7} (blue
stars for BeppoSAX and magenta circles for EP), excluding sources like
stellar flares, leads to the following conclusions:
\begin{enumerate}
\def\labelenumi{\arabic{enumi}.}
\item
the rate of the events in the two instruments is comparable when one
considers the region above the BeppoSAX-WFC sensitivity;
\item
in both instruments, particularly EP, there appears to be a lack of
events of duration shorter than about 40-50 seconds compared to longer
durations. This is consistent with the extensive evidence that the
X-ray emission in GRBs is longer than in gamma-rays
\citep[][]{piro1998};
\item
at duration larger than about 50 seconds, where the bulk of events
lies, the better EP sensitivity discloses a population of faint
events;
\item
the distribution of the faint EP events seems to connect
without gaps with brighter events (both by BeppoSAX and EP).
\end{enumerate}
The latter point suggests that the bulk of the population of faint EP
event could represent the extension of the brighter one and thus could
share the same origin, i.e. be associated to GRB-like progenitors. In
fact, in a few cases, these faint events have been indeed explained in
this context
\citep[][]{rastinejad2025,jiang2025,srinivasaragavan2025,connor2025}.
In this regard it should be pointed out that the association of this
faint population to GRBs is much more difficult because any gamma-ray
signal associated to such faint events would be undetectable by GRB
monitors currently flying. It could only be achieved by follow-up
campaigns disclosing the associated afterglows and counterparts.

Another interesting comparison to make is with MAXI. According to the
MAXI alert messages disseminated through email, MAXI has detected 150
stellar flares in 15 years from 29 unique stars which translates to a
yearly all-sky rate of approximately 15 yr$^{-1}$ above a sensitivity
threshold of about 6$\times10^{-10}$~\ecs\ \citep[2-30
  keV;][]{matsuoka2009}. This is substantially lower than WFC, WXT and
other missions. Parts of this can be explained by the worse
sensitivity and an incomplete count on our part from the
literature. MAXI has a large sky coverage for events that last longer
than one MAXI orbit and should be quite efficient in detecting these.

The census of late-type main sequence stars is complete for distances
of up to about 200 pc from GAIA measurements, for a limiting G
magnitude of 21 \citep{gaia2023}. \cite{pye2015} determines an X-ray
peak flux range for stellar flares of $10^{27}$-$10^{32}$~\lum. The
WFC sensitivity limit of 2$\times10^{-10}$~\ecs\ implies that even for
the smallest distance of 1 pc, not the whole range of peak fluxes of
stellar flares can be detected. Thus, the reach of WFC detections is
smaller than that of optical detections of late-type stars. Therefore,
any WFC detection of a stellar flare should be identifiable through an
optical counterpart and this is indeed the case.

We identified 33 stellar flares by cross checking positions with
catalogs of stars \citep[e.g.,][]{kowalski2024}. The location of an
additional event is coincident with ROSAT source 1RXS J214641.1-854303
and M3.5 star NLTT 51688, a probable flare star. Four further flares
were detected from sky regions that are not known to host a known
flare star. These have been tentatively identified with other
counterparts, one which exhibited gamma-ray flares in the 1990s (GR
Dra) and one with a cool star (UCAC4 255-003783).  Therefore, 87\% of
the long-duration/soft events could be unambiguously identified
through obvious counterparts.

42\% of the FXTs reported here have not been reported previously. This
applies to 37 out of 38 stellar flares, 3 SFXTs and 23 out of 100
GRBs. This is unfortunate for this revolutionizing phase of GRB
research when the distance scale of GRBs was first established.  The
reason is that the missed-out GRBs are mostly gamma-ray poor. This
prevented quick-look recognition of GRBs in the GRBM and quick
follow-up in WFC data for localization, specially in the early phase
of the mission lifetime. However, some were recognized in real time
and triggered broad band follow-up observations. It was then realized
that they were at similar distances as GRBs and exhibited broad-band
afterglows not unlike GRB, motivating the association of XRF to a
class of GRBs \citep[e.g.,][]{piro2005,galli2006,amati2004}. This
association has been further consolidated following the launch of HETE
II \citep{lamb2004} and Swift \citep{gehrels2009}, with XRF sample
studies of the joint BeppoSAX-HETE-2 sample
\citep{dalessio2006,sakamoto2005} and Swift \citep{sakamoto2008}

If we apply a peak flux softness ratio (see column 17 in
Table~\ref{table1}) threshold of 16 or larger, above which we fail to
detect a GRBM signal for most GRBs, for XRFs we find 37 such cases or
37\% of all GRBs. This is roughly consistent with earlier findings,
although those make a distinction between 3 classes rather than 2 and
apply thresholds to different definitions of spectral hardness ratios:
X-ray flashes (H$_{\rm s}>1$ with H$_{\rm s}$=fluence(2-30
keV)/fluence(40-400 keV), X-ray rich GRBs (0.32\textless H$_{\rm
  s}$\textless1) and classical GRBs (H$_{\rm s}$\textless0.32)
(\citet{dalessio2006}, see also \citet{sakamoto2005}). Note that the
hardest GRBs in our sample have a softness ratio of 0.8 (GRB990123 and
GRB000210). These rates are slightly different compared to the sample
detected by EP-WXT. For example, \cite{zhang2025} finds that 22\% of
FXTs detected by EP-WXT have a gamma-ray counterpart, searching in
GECAM and Fermi/GBM data. This difference likely arises from the
better sensitivity of EP-WXT, which can detect transients as faint as
a few times 10$^{-11}$~\ecs, an equivalent sensitivity that cannot be
achieved by any GRB detector currently in operation in gamma rays. The
difference in the rates could also be due, in part, to the different
bandpasses of the GRBM, with respect to Fermi-GBM (8 keV -- 40 MeV)
and GECAM (10 keV -- 6 MeV). The GRBM sensitivity, though, is
comparable to Fermi-GBM and GECAM (order of 10$^{-8}$~\ecs).  Finally,
it is obvious that the EP sample is biased because of its lack of
direct $\gamma$-ray coverage since it does not host a $\gamma$-ray
instrument.

The science goal of the WFC, when the instrument was proposed in the
1980s, was the acceleration of data collection on FXTs to better
determine their origin. The WFC has detected about 149 FXTs in the
categories of GRBs/XRFs (100), stellar flares (38) and SFXT flares
(11). The first-time fast (within hours) and accurate (within arcmin)
localization of GRBs solved the decades-long enigmatic mystery of
their origin. The topping off with stellar flares has made the
identification possible of 99\% of all FXTs and explained their
origin. The conclusion is that the WFC instrument has met its science
goal.

\begin{acknowledgements}

We are indebted to retired SRON colleagues Gerrit Wiersma and Jaap
Schuurmans for their invaluable assistance in the data reduction of
WFC data, and to Wim Mels, Rieks Jager, Bert Brinkman and Frank van
Beek for leading the instrument development of the WFC. We thank the
anonymous referee for a helpful review. LP and GG acknowledge support
from ASI (Italian Space Agency) through contract
no. 2019-27-HH.0. This research has made use of the SIMBAD database,
operated at CDS, Strasbourg, France, NASA's Astrophysics Data System
Bibliographic Services and Jochen Greiner's compendium of GRBs
available at URL {\tt www.mpe.mpg.de/$\sim$jcg/grbgen.html}.
\end{acknowledgements}

\noindent {\footnotesize {\it Author contributions.}  Data extraction:
  JZ and CG. Data analysis: JZ, LP and GG.  Manuscript preparation:
  JZ. Manuscript review and improvement: all authors. It is noted that
  no AI tool has been employed in the data reduction analysis nor
  manuscript preparation.}

\vspace{2mm}
\noindent{\footnotesize {\it Online data} concern the time series data
  of Figs. \ref{figlc1} and \ref{figlc2}. These are available at {\tt
    https://doi.org/10.5281/zenodo.18849381}}

\bibliographystyle{aa} \bibliography{aa58641-25-references}

\setcounter{figure}{2}
\begin{figure*}
\includegraphics[width=1.88\columnwidth]{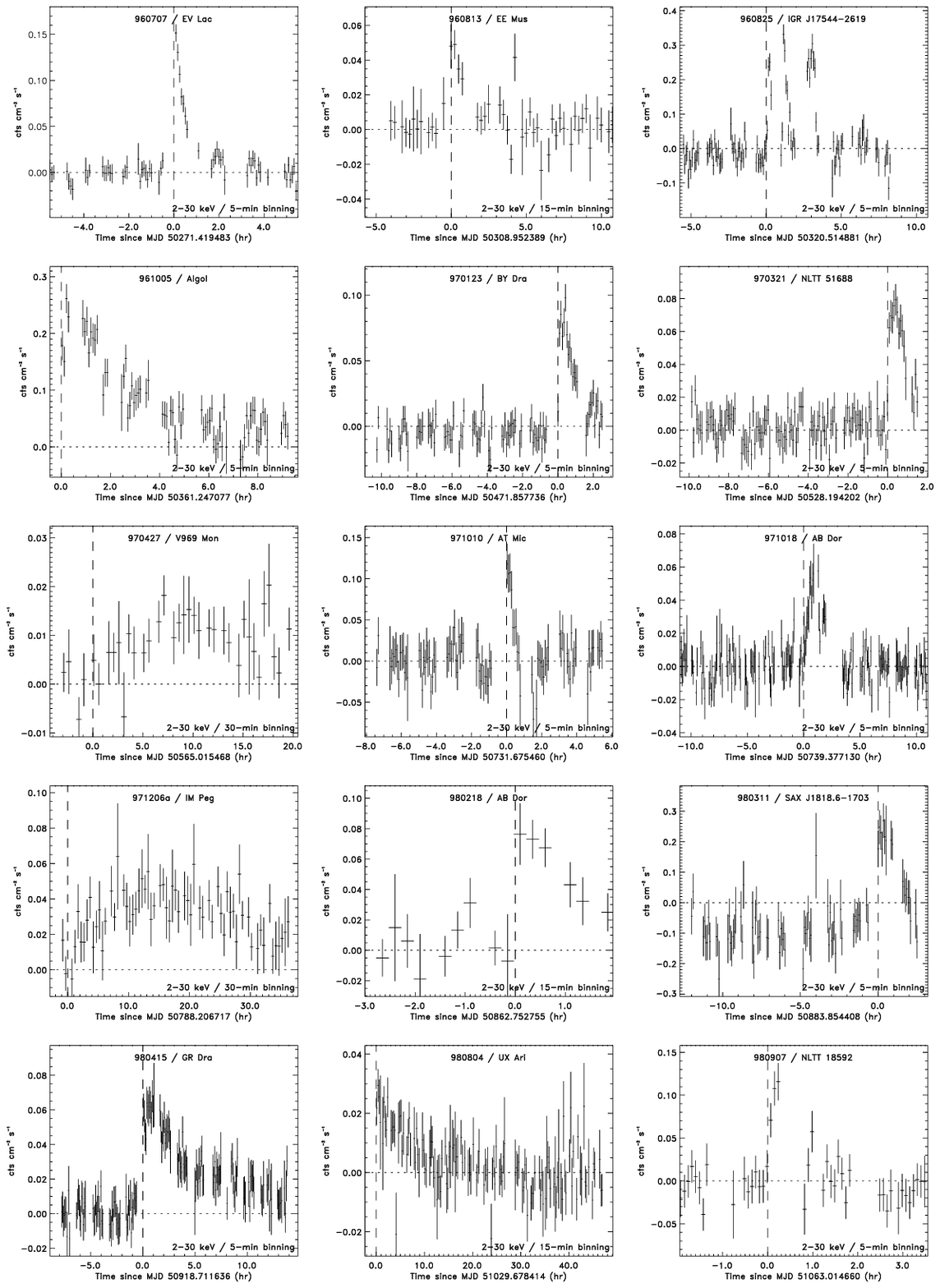}
\caption{Light curves of 49 magnetic flares.}
\label{figlc1}
\end{figure*}

\setcounter{figure}{2}
\begin{figure*}
\includegraphics[width=1.88\columnwidth]{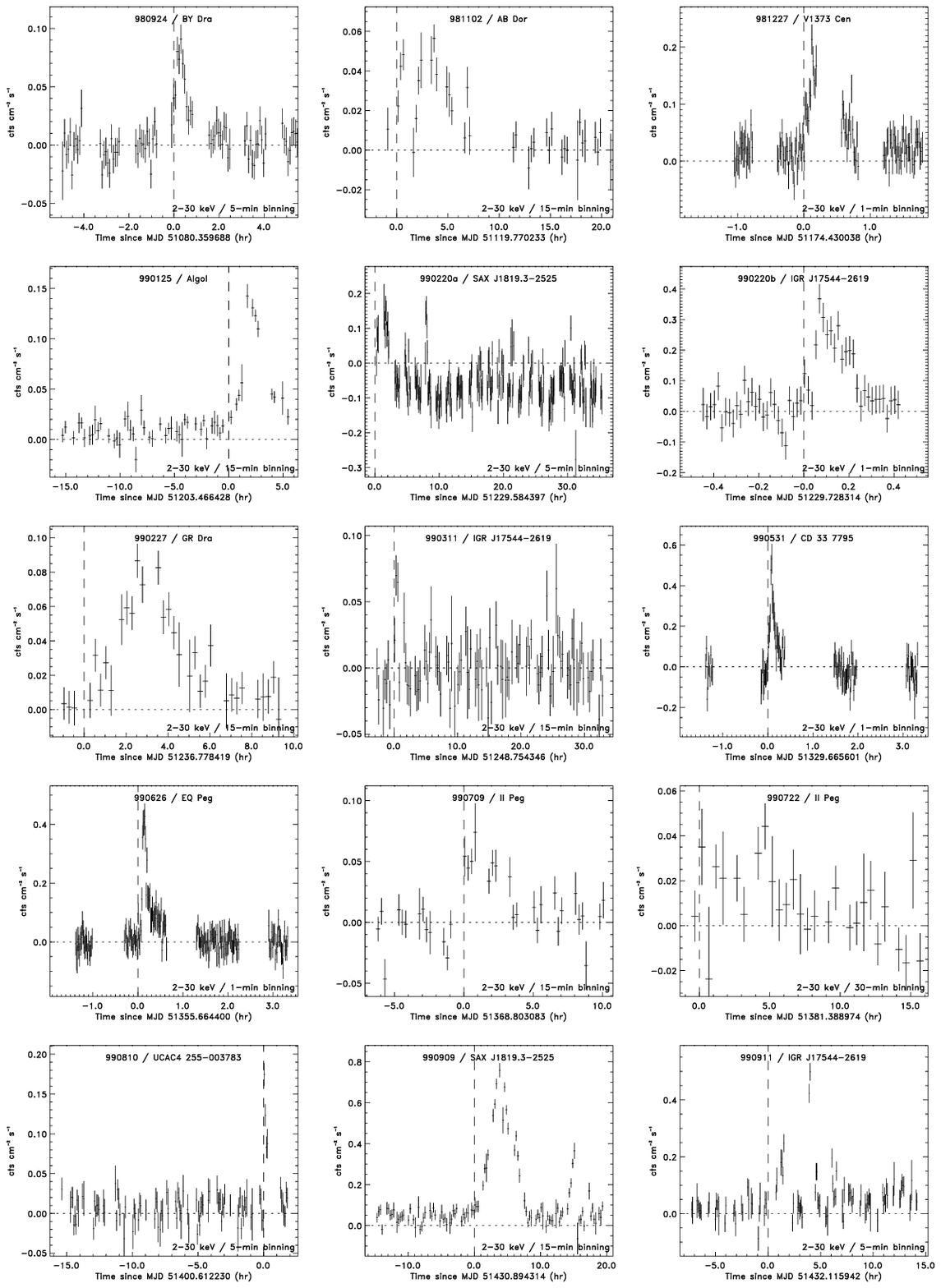}
\caption{cont'd (2/4)}
\end{figure*}

\setcounter{figure}{2}
\begin{figure*}
\includegraphics[width=1.88\columnwidth]{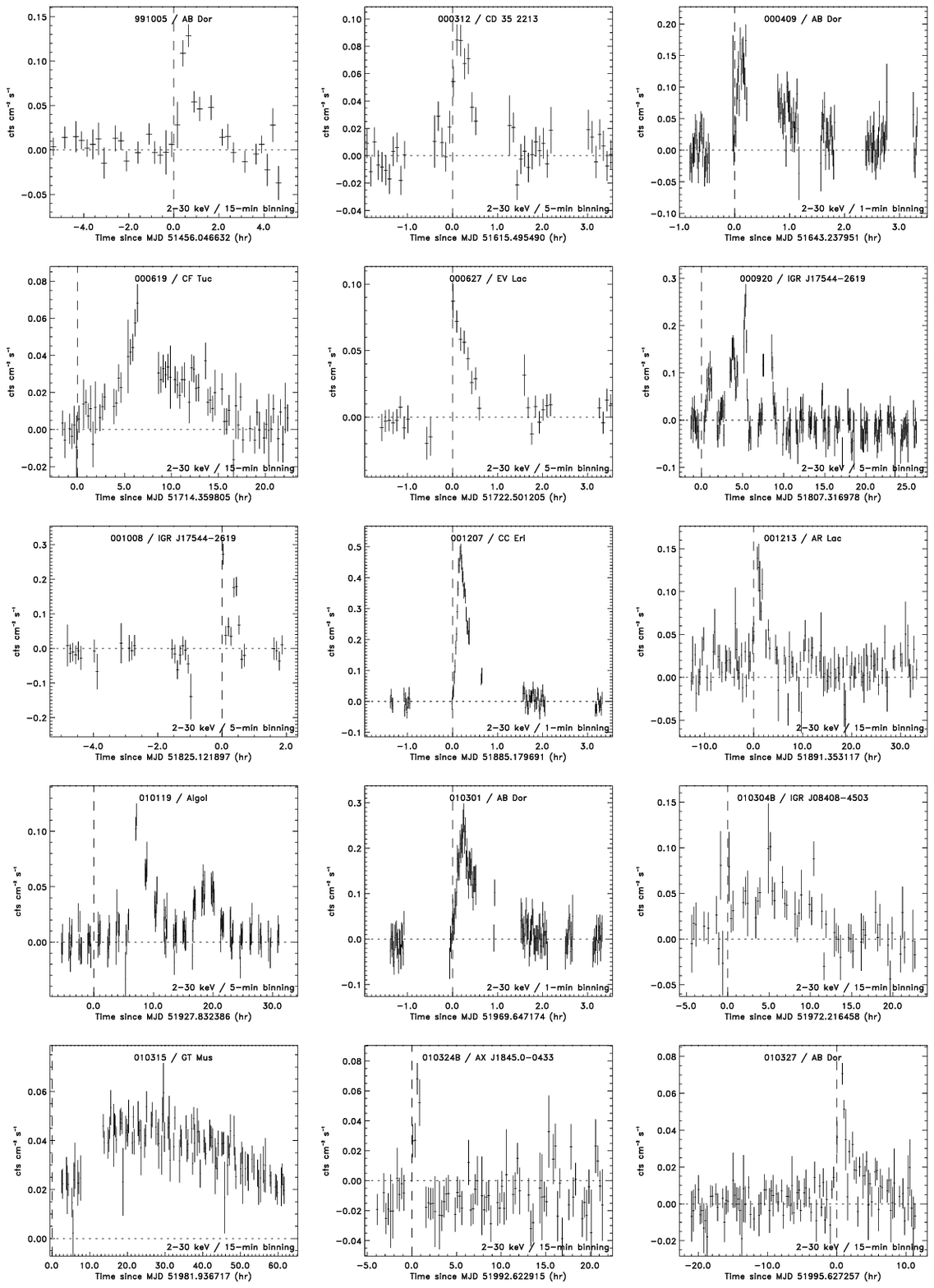}
\caption{cont'd (3/4)}
\end{figure*}

\setcounter{figure}{2}
\begin{figure*}
\includegraphics[width=1.88\columnwidth]{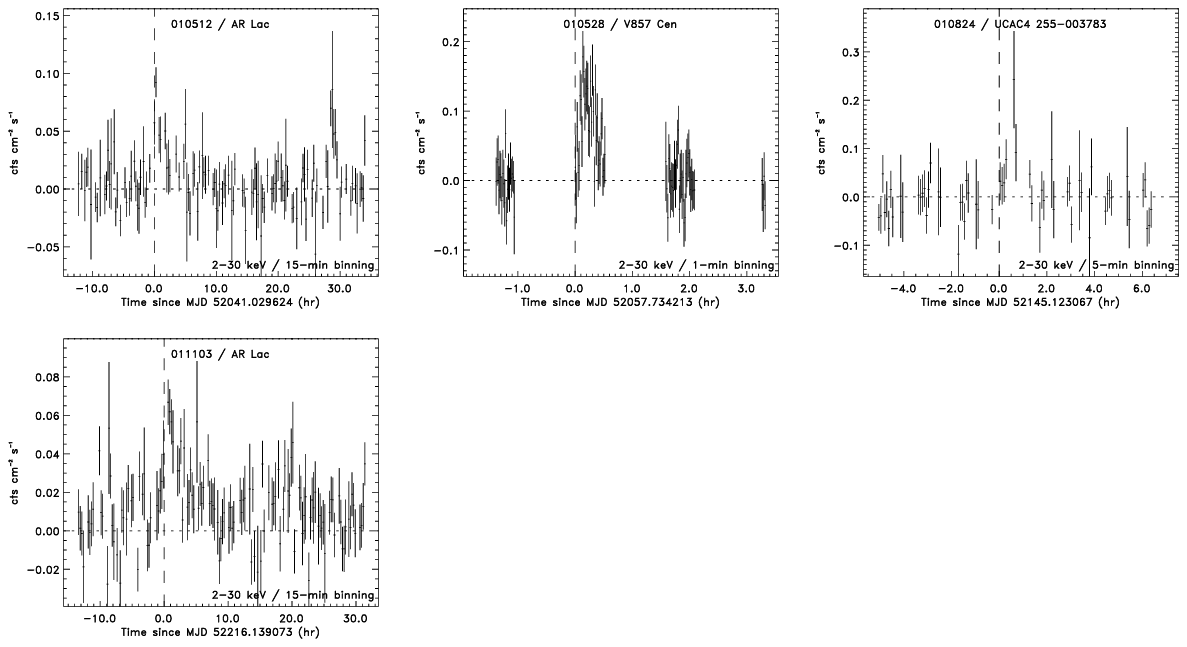}
\caption{cont'd (4/4)}
\end{figure*}

\setcounter{figure}{3}
\begin{figure*}
\includegraphics[width=1.88\columnwidth]{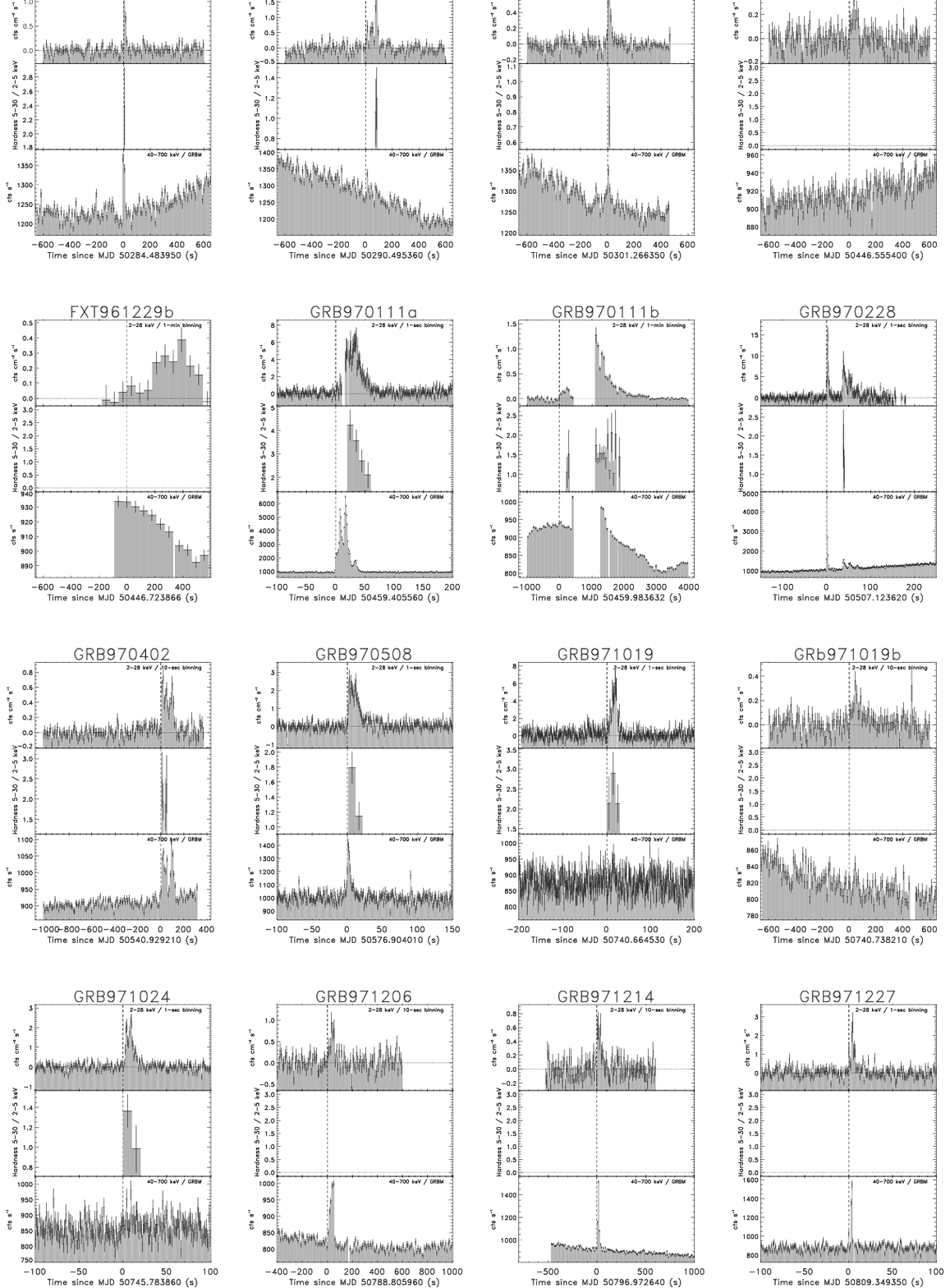}
\caption{Light curves of all 100 GRBs. Top panels: in WFC full
  bandpass (2-30 keV). Middle panels: 5-30/2-5 keV spectral hardness
  ratio from WFC data. Bottom panels: 40-700 keV GRBM light
  curves. (1/7)}
\label{figlc2}
\end{figure*}

\setcounter{figure}{3}
\begin{figure*}
\includegraphics[width=1.88\columnwidth]{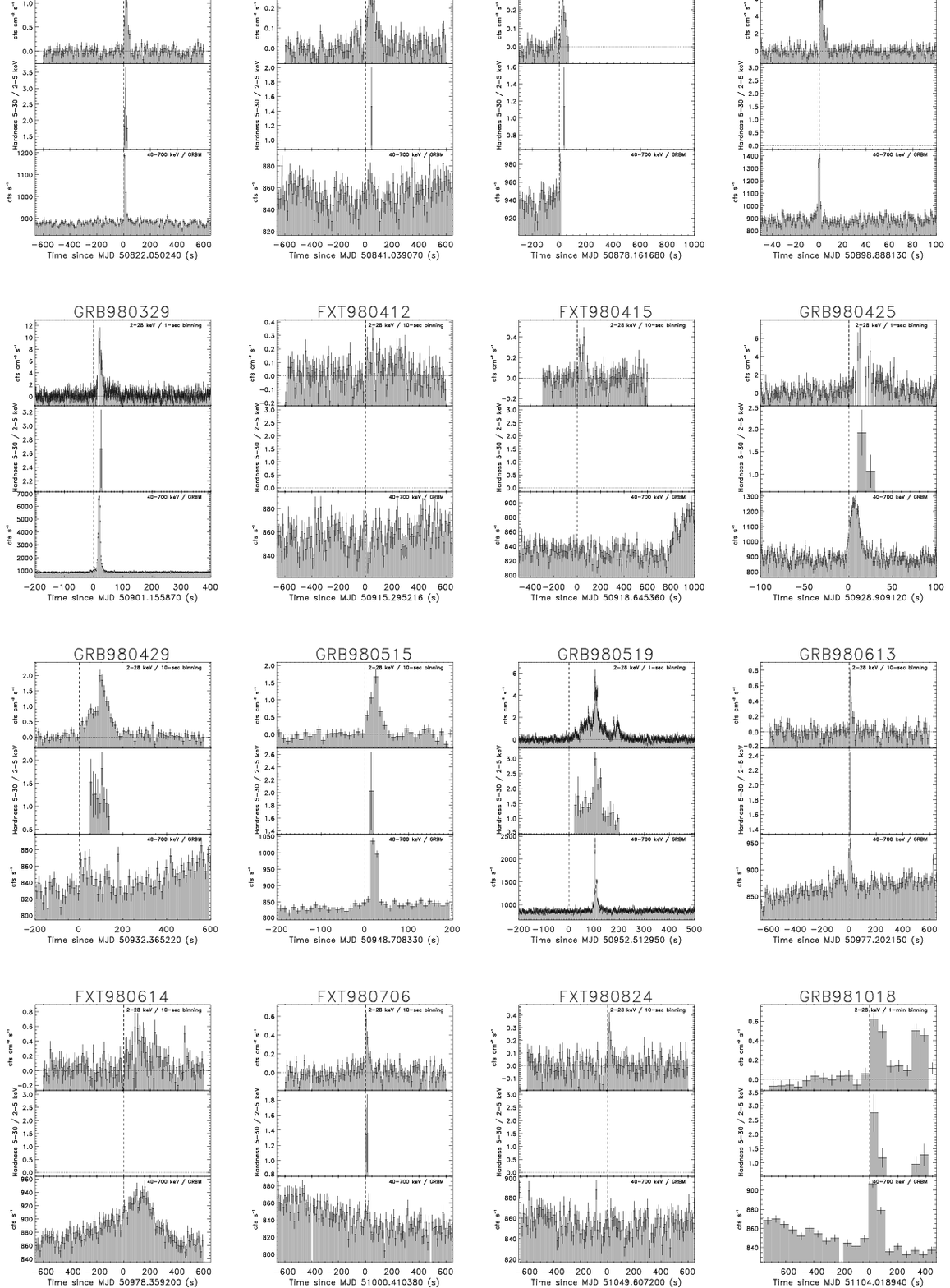}
\caption{cont'd (2/7)}
\end{figure*}

\setcounter{figure}{3}
\begin{figure*}
\includegraphics[width=1.88\columnwidth]{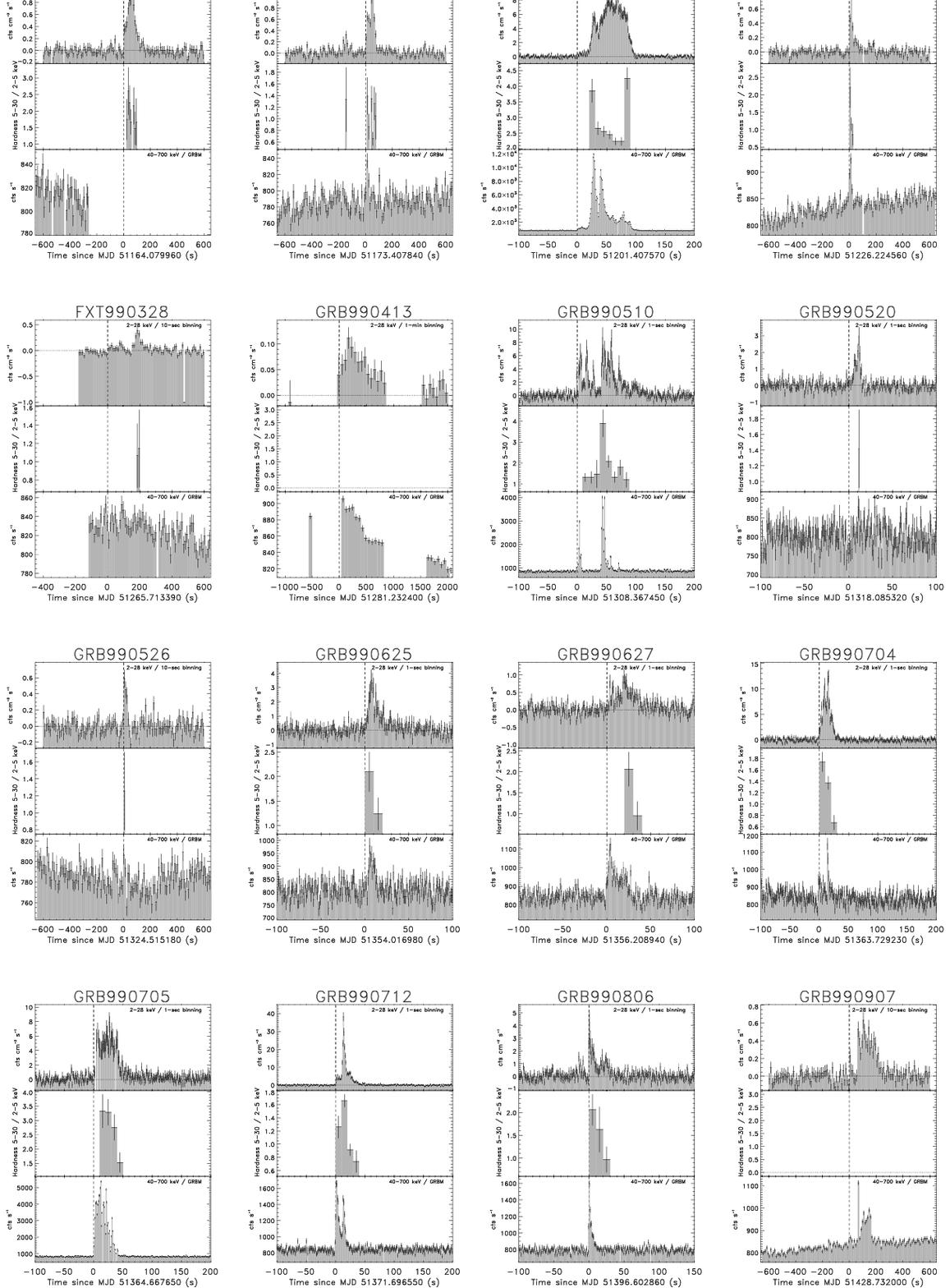}
\caption{cont'd (3/7)}
\end{figure*}

\setcounter{figure}{3}
\begin{figure*}
\includegraphics[width=1.88\columnwidth]{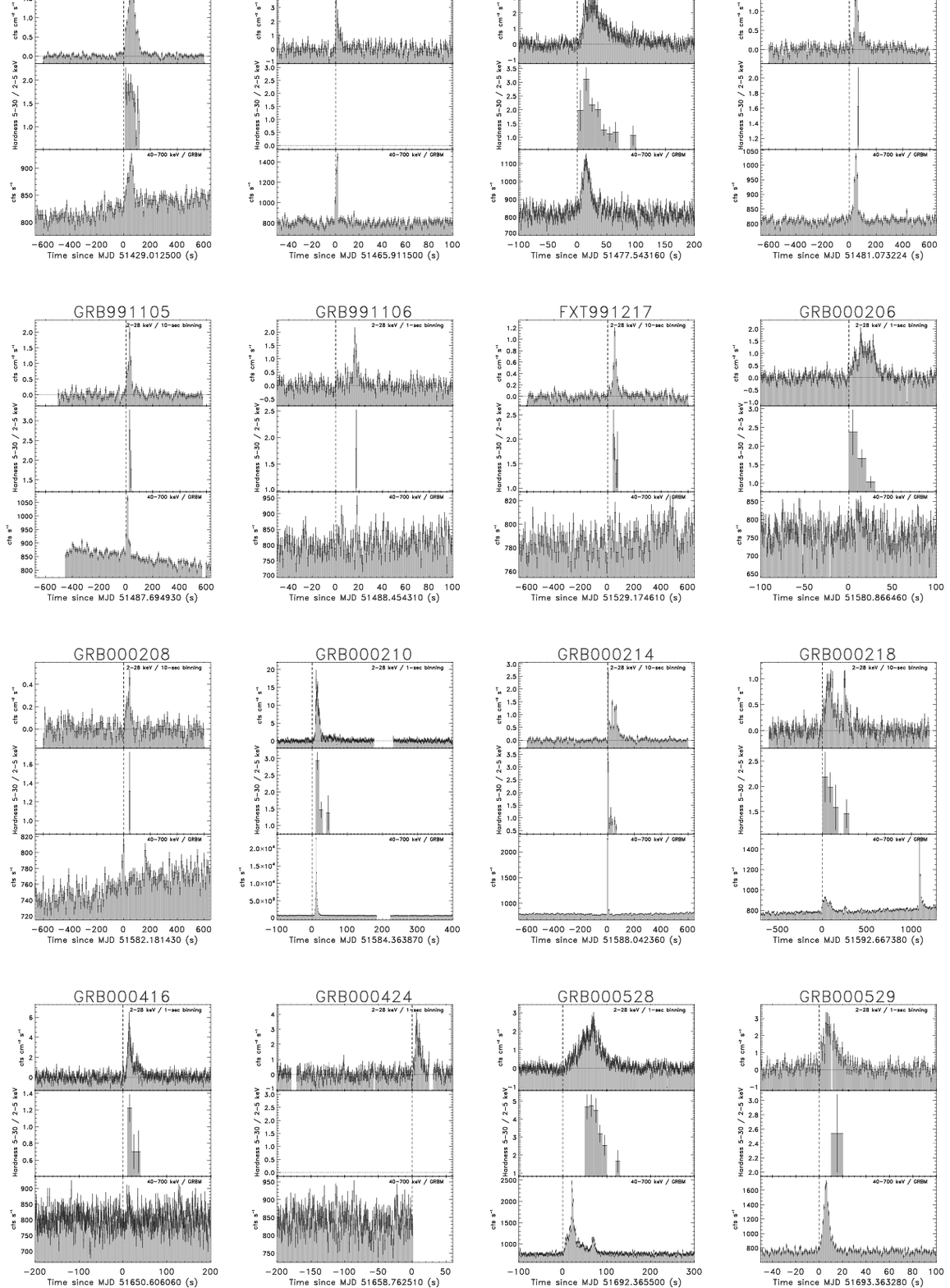}
\caption{cont'd (4/7)}
\end{figure*}

\setcounter{figure}{3}
\begin{figure*}
\includegraphics[width=1.88\columnwidth]{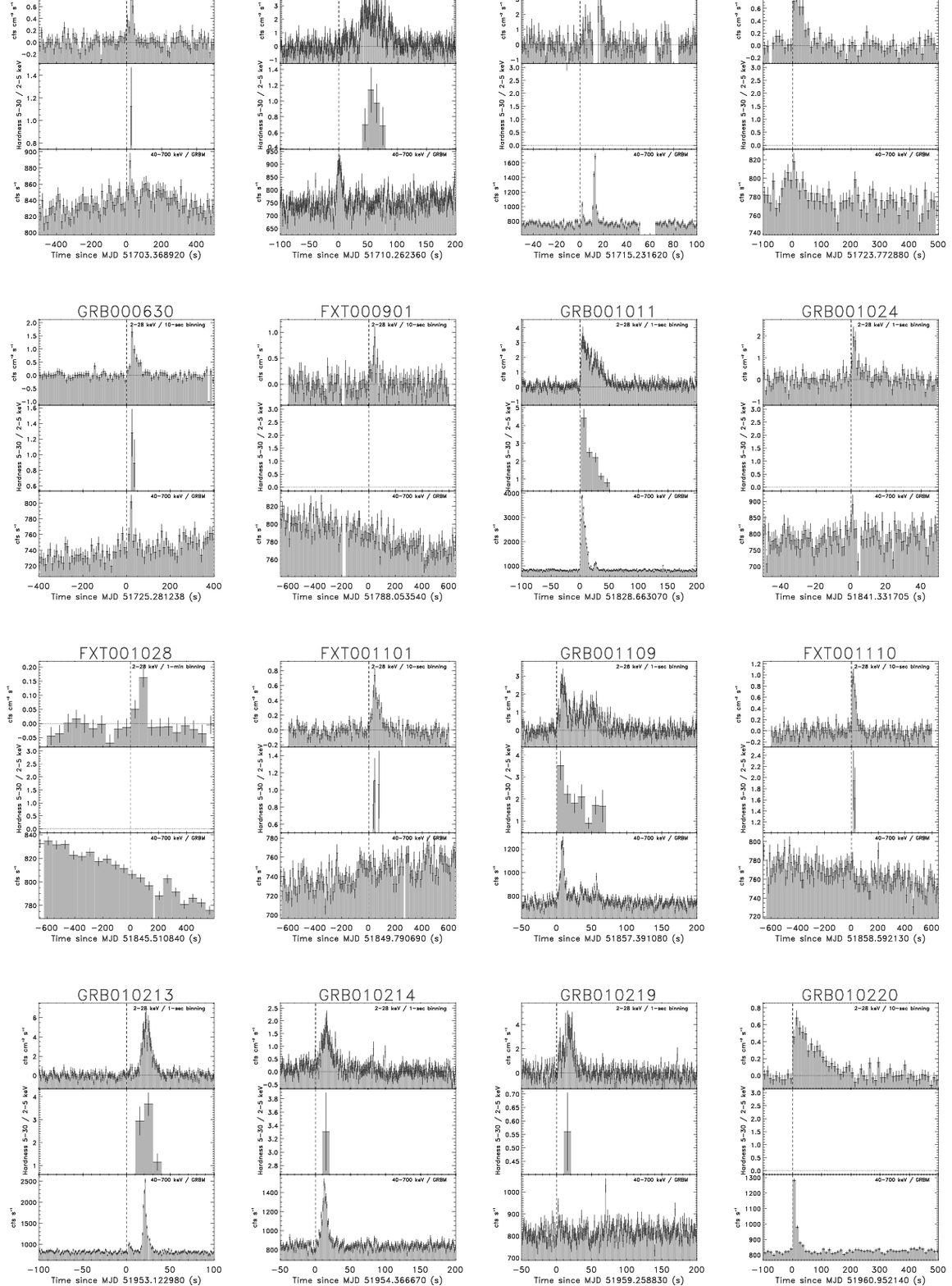}
\caption{cont'd (5/7)}
\end{figure*}

\setcounter{figure}{3}
\begin{figure*}
\includegraphics[width=1.88\columnwidth]{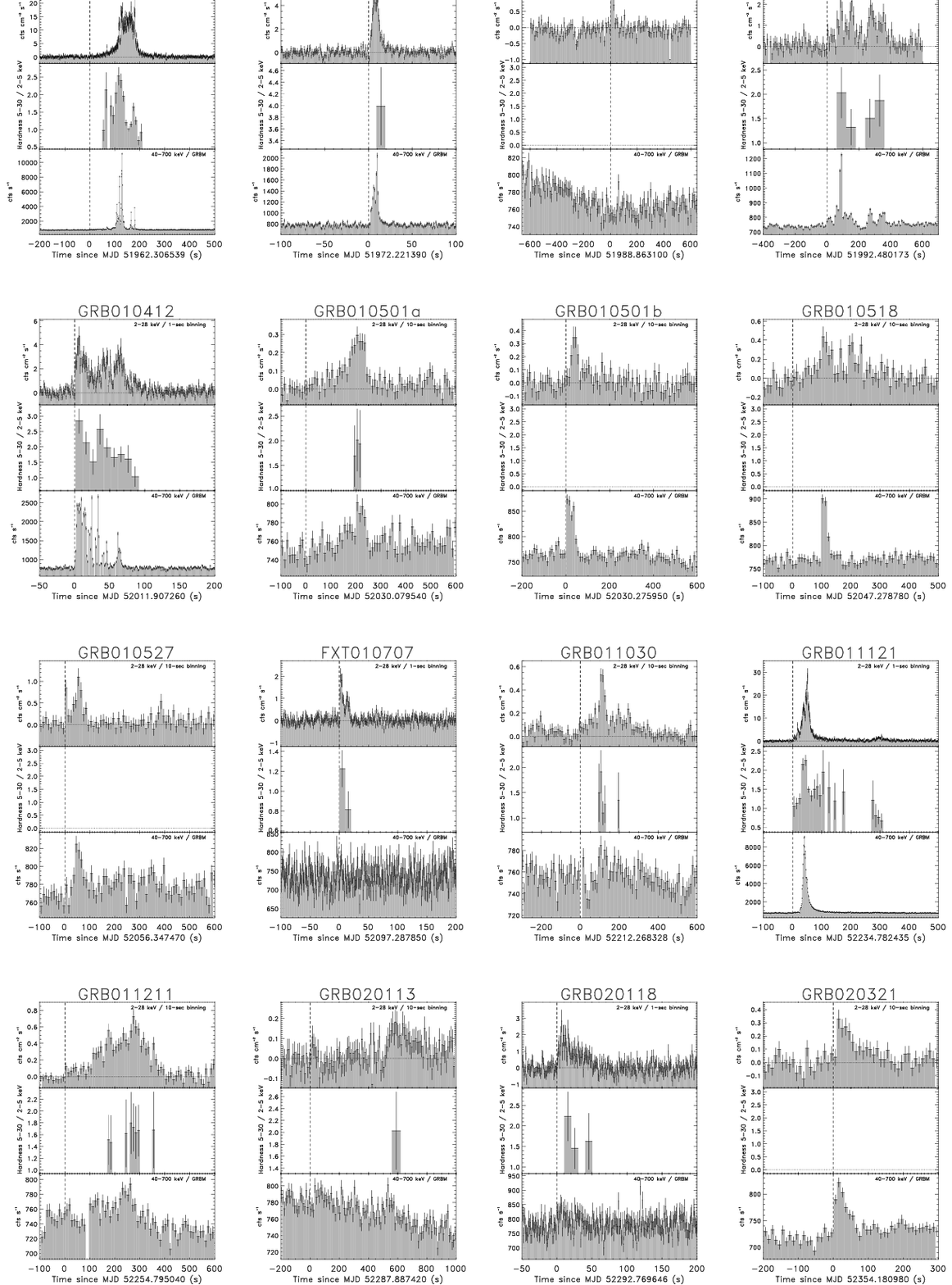}
\caption{cont'd (6/7)}
\end{figure*}

\setcounter{figure}{3}
\begin{figure*}
\includegraphics[width=1.88\columnwidth]{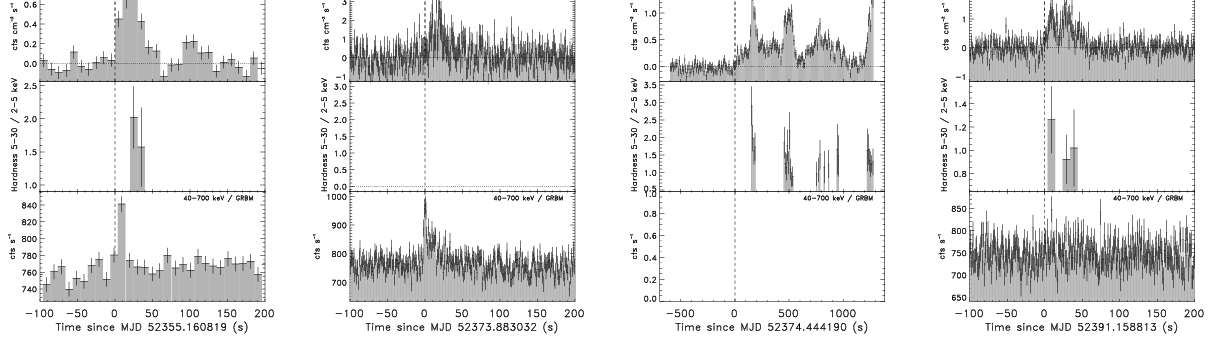}
\caption{cont'd (7/7)}
\end{figure*}

\setcounter{table}{0}
\tiny
\longtab[2]{
\begin{landscape}
\begin{longtable}{l|c|c|r|c|l|r|r|r|r|c|r|l|r|r|r|l|R}
\caption{Catalog of 149 FXTs detected with BeppoSAX-WFC. For an
explanation of this table, see Sect.~\ref{cata} of main text.}
\label{table1}
\tabularnewline
\hline\hline
ID      &Ref$^\dag$ & r & OP&WFC & MJD(UTC) &  T90 (s)&  T90 &  R.A. & Dec. & Err.        & Dev.       & BAT-& \multicolumn{2}{c|}{Peak flux}    & Soft & Type & Identification / remarks\\
        &           &   &  &     & start    &         &  err &       &      &($\arcmin$)  & ($\arcmin$)&SE  & WFC (cts          & GRBM          & ness  &      &     \\
        &           &   &  &     &          &         &  (s) &       &      &             &            &ID  & s$^{-1}$cm$^{-2}$) & (cts\,s$^{-1}$) & ratio &      &     \\
\hline
\endfirsthead
\caption{continued.}\\
ID      &Ref$^\dag$ & r & OP&WFC & MJD(UTC) & T90 (s) &  T90 &  R.A. & Dec. & Err.        & Dev.       & BAT- & \multicolumn{2}{c|}{Peak flux}    & Soft & Type & Identification / remarks\\
        &           &   &  &     & start    &         &  err &       &      & ($\arcmin$) & ($\arcmin$)&SE    & WFC (cts\,        & GRBM          & ness  &      &     \\
        &           &   &  &     &          &         &  (s) &       &      &             &            &ID    & s$^{-1}$cm$^{-2}$)& (cts\,s$^{-1}$) & ratio &      &     \\
\hline\hline
\endhead
\hline
\endfoot
960707  & u  & &  616&1 & 50271.419483 &    11728.6 &  1787.2  &  341.7196 & +44.3296 &  1.68 & 0.64 &      & 0.152 &       &       & Flare star & EV Lac flare 1 / M-dwarf flare star with superflares (ref43) \tabularnewline 
960720  & 1  &x&  675&1 & 50284.483950 &       21.1 &     8.2  &  262.6197 & +49.0970 &  2.64 &      & 5545 & 1.21  & 1220  & 1.0   & GRB & GRB970720 / First WFC GRB published, too large error reg for follow up \tabularnewline 
960726  & u  & &  689&2 & 50290.495360 &       77.7 &     2.2  &  208.8450 & -52.4484 &  2.74 &      &      & 1.81  &  190  & 9.8   & GRB & GRB960726 \tabularnewline 
960806  & u  & &  735&1 & 50301.266350 &       27.1 &     4.8  &   32.6648 &  -0.6359 &  2.91 &      &      & 0.65  &  155  & 4.3   & GRB & GRB960806 \\
960813  & u  & &  757&2 & 50308.952389 &    15368.9 &   520.6  &  187.8133 & -65.8235 &  3.15 & 2.49 &      & 0.049 &       &       & Flare star & EE Mus=1RXS J123052.5-655020 \\
960825  & 2  & &  820&2 & 50320.514881 &    11021.7 &   442.4  &  268.6070 & -26.3305 &  1.70 & 0.27 &      & 0.332 &       &       & SFXT & IGR J17544-2619 flare 1 / SFXT \\
961005  & u  & & 1075&2 & 50361.247077 &    19486.5 &   816.1  &   47.0574 & +40.9467 &  1.67 & 0.88 &      & 0.262 &       &       & Flare star & Algol flare 1 / Flare star in a binary with an early-type star \\
961229a & u  & & 1448&2 & 50446.555400 &       52.9 &     7.9  &  224.7979 & +18.5786 &  6.02 &      &      & 0.36  & $<$20 & $>$18 & GRB? & FXT961229a / XRF coincident with BD +19 2895 (K0 star)? \\
961229b & 3  & & 1448&1 & 50446.723866 &      339.3 &    41.0  &   38.7055 &  -0.8724 &  3.31 &      &      & 0.39  & $<$20 & $>$20 & GRB? & FXT961229b / XRF coincident with PKS J0234-0049 (RADIO GALAXY) \\
970111a & 4  &+& 1485&2 & 50459.405560 &       43.3 &     2.3  &  232.0354 & +19.5933 &  1.74 &      &      & 6.60  & 5800  & 1.1   & GRB & GRB970111 / First GRB with follow up \\
970111b & 3  & & 1487&1 & 50459.983632 &     1567.0 &   148.9  &   86.1699 & -40.7063 &  1.56 &      &      & 1.27  & 90    & 14    & GRB & GRB970111b / Long GRB \\
970123  & u  & & 1565&2 & 50471.857736 &     6999.3 &   370.9  &  278.4714 & +51.7236 &  1.91 & 0.26 &      & 0.098 &       &       & Flare star & BY Dra flare 1 / M-dwarf flare star \\
970228  & 5  &+& 1671&1 & 50507.123620 &       68.5 &     1.2  &   75.4315 & +11.7859 &  1.78 & 0.81 &      & 15.01 &  3900 & 3.8   & GRB & GRB970228 / First GRB with successful follow up and distance \\
970321  & u  & & 1780&1 & 50528.194202 &     4634.3 &   189.3  &  326.4788 & -85.7153 &  1.73 & 0.87 &      & 0.079 &       &       & Flare star? & NLTT 51688=1RXS J214641.1-854303 / M3.5 star with no known flare activity \\
970402  & 6  &+& 1830&1 & 50540.928920 &      103.6 &     4.4  &  222.6150 & -69.3224 &  4.66 &      &      & 0.75  &   160 & 4.7   & GRB & GRB970402 \\
970427  & u  & & 1930&2 & 50565.015468 &    57767.6 &  3532.6  &   99.2513 &  -5.3676 &  2.82 & 1.33 &      & 0.018 &       &       & RS CVn & V969 Mon=1RXS J063656.7-052104 / Faintest detected flare \\
970508  & 7  &+& 1984&2 & 50576.904030 &       40.7 &     2.2  &  103.4250 & +79.2696 &  1.94 &      & 6225 & 2.71  &   410 & 6.6   & GRB & GRB970508 / First GRB with radio scintillation and evidence for beaming \\
971010  & u  & & 2622&2 & 50731.675460 &     1731.6 &   464.3  &  310.4521 & -32.4547 &  3.75 & 1.51 &      & 0.117 &       &       & Flare star & AT Mic / dwarf M flare star \\
971018  & u  & & 2699&2 & 50739.377130 &     9869.6 &  1541.3  &   82.1835 & -65.4403 &  2.22 & 0.62 &      & 0.057 &       &       & T Tau star & AB Dor flare 1 \\
971019a & 8  & & 2705&2 & 50740.664410 &       22.2 &     1.9  &  217.6194 & +74.3645 &  2.56 &      & untr & 6.09  &    55 & 111   & GRB & GRB971019a / XRF \\
971019b & u  & & 2706&1 & 50740.738210 &      446.8 &    33.6  &   88.9121 & -74.9495 &  3.61 &      &      & 0.36  &    35 & 10    & GRB & GRB971019b \\
971024  & a  & & 2732&1 & 50745.783860 &       16.1 &     1.5  &  140.6739 & +51.3087 &  2.04 &      & untr & 2.46  &    60 & 41    & GRB & GRB971024 / XRR \\
971206a & u  & & 3066&2 & 50788.206717 &   105662.9 &  2871.4  &  343.2672 & +16.8327 &  1.89 & 0.46 &      & 0.042 &       &       & RS CVn & IM Peg \\
971206b & 9  & & 3066&1 & 50788.805960 &       76.3 &     9.6  &  161.0457 &  -5.7792 &  4.44 &      & 6521 & 0.94  &   180 & 5.2   & GRB & GRB971206 \\
971214  & 10 &+& 3123&1 & 50796.972640 &       86.1 &   22.8   &  179.1035 & +65.1916 &  4.06 & 0.54 & 6533 & 0.64  &   585 & 1.1   & GRB & GRB971214 \\
971227  & 11 &+& 3205&2 & 50809.349350 &       14.6 &     1.2  &  194.4174 & +59.2444 &  3.70 &      & 6546 & 2.69  &   650 & 4.1   & GRB & GRB971227 \\
980109  & 12 &x& 3289&2 & 50822.050240 &       36.4 &    18.6  &    6.5210 & -63.0221 &  2.29 &      & 6564 & 1.25  &   285 & 4.4   & GRB & GRB980109 / No afterglow found (no NFI) \\
980128  & 8  & & 3465&2 & 50841.039070 &      112.4 &    24.5  &   15.9980 & +58.5142 &  2.78 &      & untr & 0.26  & $<$20 & $>$13 & GRB & GRB980128 / XRR \\
980218  & u  & & 3714&1 & 50862.752755 &     5561.7 &   453.1  &   82.1791 & -65.4693 &  3.09 & 1.13 &      & 0.076 &       &       & T Tau star & AB Dor flare 2 \\
980306  & a  & & 3875&2 & 50878.161680 &       38.3 &     2.7  &  240.8666 & -77.8760 &  2.91 &      & untr & 0.30  &    35 & 8.4   & GRB & GRB980306 / XRR \\
980311  & 13 & & 3905&2 & 50883.855408 &     6861.6 &   630.7  &  274.7542 & -16.9006 &  5.00 & 10.57&      & 0.269 &       &       & SFXT & SAX J1818.6-1703 \\
980326  & 14 &x& 4057&1 & 50898.888130 &        5.8 &     2.4  &  129.1306 & -18.8800 &  3.19 & 1.56 & 6660 & 6.22  &   480 & 13    & GRB & GRB980326 \\
980329  & 15 &+& 4073&2 & 50901.155870 &       64.4 &     3.7  &  105.6408 & +38.8432 &  3.22 & 0.84 & 6665 & 9.29  &  6345 & 1.5   & GRB & GRB980329 \\
980412  & u  & & 4169&2 & 50915.295216 &      480.4 &    24.6  &  123.0199 & -49.5219 &  4.74 &      &      & 0.27  & $<$20 & $>$14 & GRB? & FXT980412 / XRF, coincides with HD 6881 (K3(III) star) \\
980415  & u  & & 4202&2 & 50918.645360 &       73.3 &     8.4  &  246.6110 & +63.5242 &  5.00 &      &      & 0.29  & $<$20 & $>$14 & GRB? & FXT980415 / XRF \\
980415b & u  & & 4202&2 & 50918.711636 &    41707.5 &  1623.7  &  262.1440 & +59.0247 &  1.64 & 0.89 &      & 0.061 &       &       & \parbox[t]{1cm}{High-E \\transient?} & GR Dra flare 1 / See also 990227 \\
980425  & 16 &+& 4281&2 & 50928.909030 &       49.7 &     3.6  &  293.7303 & -52.8553 &  2.63 & 1.31 & 6707 & 6.34  &   400 & 16    & GRB & GRB980425=SN1998bw / first supernova Ic association \\
980429  & 17 & & 4320&2 & 50932.365870 &      129.8 &     5.8  &  130.1491 & +22.8360 &  1.92 &      &      & 1.99  &    35 & 57    & GRB & GRB980429 / XRF \\
980515  & 18 &+& 4462&2 & 50948.708330 &       30.8 &     3.7  &  319.3480 & -67.2500 &  2.63 &      &      & 1.67  &   185 & 9.0   & GRB & GRB980515 / no counterpart found in other wavelengths than X-rays \\
980519  & 19 &+& 4480&2 & 50952.512600 &      160.8 &     3.4  &  350.5923 & +77.2505 &  1.53 & 0.69 & 6764 & 6.15  &  1515 & 4.1   & GRB & GRB980519 / very soft prompt black body X-ray spectrum \\
980613  & 9  &+& 4677&2 & 50977.202150 &       53.0 &     9.5  &  154.4848 & +71.4853 &  3.42 & 1.68 &      & 0.71  &    85 & 8.3   & GRB & GRB980613 \\
980614  & u  & & 4683&1 & 50978.359433 &      301.2 &    16.2  &  337.7366 & +33.3625 &  5.00 &      &      & 0.57  &    20 & $>$29 & GRB? & FXT980614 \\
980706  & u  & & 4894&2 & 51000.410380 &       83.7 &    12.7  &  193.1647 & +28.0604 &  3.19 &      &      & 0.57  & $<$20 & $>$38 & GRB? & FXT980706 \\
980804  & u  & & 5072&2 & 51029.678414 &    71073.1 &  4338.6  &   51.6457 & +28.6986 &  2.39 & 1.07 &      & 0.029 &       &       & RS CVn & UX Ari / Very long stellar flare \\
980824  & u  & & 5166&2 & 51049.607200 &       46.4 &    16.1  &  268.3526 & +48.5036 &  4.24 &      &      & 0.34  & $<$20 & $>$17 & GRB? & FXT980824 / XRF \\
980907  & u  & & 5275&1 & 51063.014660 &     6656.2 &   554.0  &  117.2305 & -76.6945 &  2.45 & 1.13 &      & 0.116 &       &       & Flare star & NLTT18592=2RE J074922-764149 / dMe star; 555 TESS flares (ref45) \\
980924  & u  & & 5385&1 & 51080.359688 &     6910.1 &   606.1  &  278.5036 & +51.7251 &  2.35 & 0.94 &      & 0.091 &       &       & Flare star & BY Dra flare 2 / M-dwarf flare star \\
981018  & 3  & & 5618&2 & 51104.017550 &      356.1 &     7.6  &  110.2276 &  +1.0364 &  2.11 &      & untr & 0.62  &    35 & 18    & GRB & GRB981018 \\
981102  & u  & & 5745&1 & 51119.770233 &    23037.3 &  2736.7  &   82.1867 & -65.4456 &  2.20 & 0.30 &      & 0.057 &       &       & T Tau star & AB Dor flare 3 \\
981217  & 9  & & 6002&2 & 51164.079960 &      111.0 &    10.7  &  294.6259 & +80.5517 &  1.95 &      &      & 0.98  &  n.a. &       & GRB? & FXT981217 \\
981226  & 20 &+& 6055&1 & 51173.407840 &      136.1 &    18.3  &  352.4125 & -23.9166 &  1.81 &      &      & 1.12  &    60 & 19    & GRB & GRB981226 / XRF \\
981227  & u  & & 6069&2 & 51174.430038 &     2365.3 &   202.2  &  175.1515 & -62.0102 &  2.18 & 1.37 &      & 0.213 &    50 & 4.3   & Flare star & V1373 Cen=1RXS J114028.6-620142 / BY Dra type star \\
990123  & 21 &+& 6221&1 & 51201.407570 &       61.0 &     0.4  &  231.3719 & +44.7483 &  1.44 & 1.11 & 7343 & 9.39  & 11200 & 0.8   & GRB & GRB990123 / 2nd brightest in GRBM \\
990125  & u  & & 6231&2 & 51203.466428 &    12755.2 &  1190.7  &   47.0381 & +40.9607 &  1.64 & 0.36 &      & 0.148 &       &       & Flare star & Algol 2 / flaring K star in a binary with an early-type star (ref44) \\
990217  & 9  &+& 6383&2 & 51226.224560 &      172.5 &    24.5  &   45.7130 & -53.0933 &  2.31 &      &      & 0.86  &    85 & 10    & GRB & GRB990217 \\
990220a & 22 & & 6419&1 & 51229.584397 &    27016.4 &   736.8  &  274.8431 & -25.4646 &  4.23 & 3.84 &      & 0.161 &       &       & HMXB & SAX J1819.3-2525 flare 3 \\
990220b & 23 & & 6419&1 & 51229.728314 &      618.5 &    65.4  &  268.6071 & -26.3230 &  2.13 & 0.19 &      & 0.368 &       &       & SFXT & IGR J17544-2619 flare 2 \\
990227  & u  & & 6473&2 & 51236.778419 &    18847.8 &  1888.9  &  262.1621 & +59.0221 &  1.93 & 0.79 &      & 0.083 &       &       & \parbox[t]{1cm}{High-E \\transient?} & GR Dra flare 2 / G0 star unknown as flare star, possibly GRO J1753+57 (gamma-ray flarer). See also 980415b \\
990311  & 23 & & 6570&1 & 51248.754346 &     5152.2 &   950.8  &  268.5934 & -26.3139 &  3.80 & 1.04 &      & 0.070 &       &       & SFXT & IGR J17544-2619 flare 6 \\
990328  & u  & & 6742&1 & 51265.715390 &      296.1 &    26.0  &  136.5108 & -54.9291 &  2.84 &      &      & 0.39  & $<$20 & $>$19 & GRB? & FXT990328 / XRF \\
990413  & 3  & & 6829&2 & 51281.232860 &     1735.4 &   138.2  &  160.1040 & +81.8236 &  2.54 &      &      & 0.11  &    60 & 1.8   & GRB & GRB990413 / long GRB \\
990510  & 24 &+& 6902&2 & 51308.367430 &      103.0 &     3.9  &  204.5196 & -80.4961 &  1.60 & 0.11 & 7560 & 8.62  &  3105 & 2.8   & GRB & GRB990510 / bright in WFC \\
990520  & 8  &+& 6952&1 & 51318.085320 &       10.4 &     1.4  &  128.9784 & +51.3096 &  2.99 &      & untr & 2.95  & $<$80 & $>$37 & GRB & GRB990520 / XRF \\
990526  & a  & & 6975&1 & 51324.515180 &       26.6 &    21.0  &  155.6031 & -62.7798 &  3.09 &      & untr & 0.59  &    25 & 24    & GRB & GRB990526 / XRF \\
990531  & u  & & 6990&1 & 51329.665601 &     2914.6 &   717.5  &  172.9817 & -34.6066 &  2.16 & 0.10 &      & 0.533 &       &       & T Tau star & CD-33 7795=1RXS J113155.7-343632 / 32 flares seen with TESS (Ref46) \\
990625  & 25 &x& 7072&2 & 51354.016980 &       36.3 &     2.8  &    6.6877 & -31.2028 &  2.68 &      &      & 3.50  &   175 & 20    & GRB & GRB990625 / too large initial error box for follow up \\
990626  & u  & & 7088&2 & 51355.664400 &     1694.3 &   397.2  &  352.9638 & +19.9369 &  1.90 & 0.27 &      & 0.427 &       &       & Flare star & EQ Peg / M-dwarf flare star with superflares \\
990627  & 26 &+& 7096&2 & 51356.208940 &       53.8 &     2.3  &   27.1347 & -77.0958 &  2.32 &      &      & 0.96  &   300 & 3.2   & GRB & GRB990627 \\
990704  & 27 &+& 7141&1 & 51363.729230 &       21.9 &     1.1  &  184.8717 &  -3.8021 &  1.90 &      &      & 12.48 &   310 & 40    & GRB & GRB990704 / XRF \\
990705  & 28 &+& 7147&2 & 51364.667650 &       69.0 &     3.8  &   77.4676 & -72.1307 &  1.78 & 0.28 &      & 6.65  &  4720 & 1.4   & GRB & GRB990705 \\
990709  & u  & & 7167&1 & 51368.803083 &    15224.4 &  1735.9  &  358.7690 & +28.6291 &  2.74 & 0.64 &      & 0.074 &       &       & RS CVn & II Peg flare 1 \\
990712  & 29 &x& 7188&2 & 51371.696550 &       29.6 &     1.0  &  337.9701 & -73.4058 &  1.50 & 0.12 &      & 38.22 &   820 & 47    & GRB & GRB990712 / Brightest WFC GRB \\
990722  & u  & & 7263&2 & 51381.388974 &    29925.5 &  3012.1  &  358.7840 & +28.6111 &  4.82 & 1.91 &      & 0.044 &       &       & RS CVn & II Peg flare 2 \\
990806  & 30 &+& 7348&1 & 51396.602510 &       34.0 &     3.4  &   47.6466 & -68.1245 &  2.13 &      & 7701 & 4.28  &   785 & 5.5   & GRB & GRB990806 \\
990810  & u  & & 7362&2 & 51400.612230 &     4678.7 &   580.9  &   56.6564 & -39.1073 &  2.17 & 0.37 &      & 0.174 &       &       & T Tauri? & SAX J0346.6-3906=UCAC4 255-003783? / This is a 16 mag 3700K star (see also 010824) \\
990907  & 9  &+& 7523&2 & 51428.732000 &      223.0 &    24.9  &  112.6887 & -69.4057 &  1.99 &      & 7755 & 0.63  &   275 & 2.3   & GRB & GRB990907 \\
990908  & 9  &x& 7523&2 & 51429.012500 &      104.3 &     5.8  &  103.4133 & -75.0104 &  1.50 &      &      & 1.76  &    90 & 20    & GRB & GRB990908 \\
990909  & 22 & & 7546&2 & 51430.894314 &    20651.4 &   759.2  &  274.8459 & -25.3962 &       &      &      & 0.760 &       &       & HXMB & SAX J1819.3-2525 flare 2 \\
990911  & u  & & 7553&1 & 51432.115942 &    27726.3 &  1690.4  &  268.6409 & -26.3300 &  1.97 & 1.83 &      & 0.499 &       &       & SFXT & IGR J17544-2619 flare 3 \\
991005  & u  & & 7692&2 & 51456.046632 &     6486.0 &   573.7  &   82.1773 & -65.4560 &  2.25 & 0.38 &      & 0.129 &       &       & T Tau star & AB Dor flare 4 \\
991014  & 31 &+& 7749&2 & 51465.911500 &        7.6 &     1.2  &  102.7685 & +11.5979 &  3.75 &      & 7803 & 3.88  &   620 & 6.3   & GRB & GRB991014 \\
991026  & 9  & & 7820&2 & 51477.543160 &      106.3 &     4.0  &  248.7926 & -89.5048 &  1.65 &      &      & 3.28  &   315 & 10    & GRB & GRB991026 \\
991030  & 9  & & 7840&2 & 51481.073455 &      114.2 &    13.7  &  122.5798 & -48.4265 &  2.40 &      &      & 1.49  &   230 & 6.5   & GRB & GRB991030 \\
991105  & 9  &x& 7859&2 & 51487.694720 &      111.4 &    27.2  &  180.8616 & -66.7735 &  2.12 &      & 7841 & 1.98  &   215 & 9.2   & GRB & GRB991105 \\
991106  & 9  &+& 7860&1 & 51488.454310 &       32.6 &     2.5  &  336.2170 & +54.3647 &  3.21 &      &      & 3.67  &   110 & 33    & GRB & GRB991106 / XRR \\
991217  & 32 &x& 8073&2 & 51529.174610 &      107.3 &    16.0  &  345.8144 &  +0.2494 &  4.5  &      &      & 1.13  & $<$20 & $>$57 & GRB? & FXT991217 \\
000206  & 33 & & 8374&1 & 51580.866460 &       39.3 &     3.1  &  226.4360 & +72.1146 &  1.86 &      & untr & 1.75  &    50 & 35    & GRB & GRB000206 / XRR \\
000208  & a  & & 8386&1 & 51582.181370 &       58.9 &    19.7  &   28.9025 & -22.8470 &  3.62 &      & untr & 0.44  &    50 & 8.9   & GRB & GRB000208 / XRR \\
000210  & 34 &+& 8406&1 & 51584.363870 &       61.8 &     3.1  &   29.8032 & -40.6742 &  1.77 & 1.05 &      & 17.17 & 20710 & 0.8   & GRB & GRB990210 / Brightest GRBM GRB \\
000214  & 35 &+& 8425&1 & 51588.042360 &      117.6 &    15.8  &  283.5719 & -66.4414 &  1.62 &      &      & 2.59  &  1825 & 1.4   & GRB & GRB000214 \\
000218  & u  & & 8468&2 & 51592.667380 &      377.1 &    35.1  &  114.4409 & -71.3064 &  1.93 &      &      & 0.98  &   185 & 5.3   & GRB & GRB000218 \\
000312  & u  & & 8683&2 & 51615.49454  &     4538.2 &   881.3  &   79.3413 & -35.3630 &  2.20 & 0.27 &      & 0.085 &       &       & Flare star & CD-35 2213=2RE J051724-352221 / dMe star). 110 flares with TESS (Ref34) \\
000409  & u  & & 8881&1 & 51643.237951 &    10243.4 &   691.1  &   82.1924 & -65.4477 &  2.05 & 0.24 &      & 0.174 &       &       & T Tau star & AB Dor flare 5 \\
000416  & 9  &x& 8942&2 & 51650.606060 &       68.4 &     4.5  &  258.8558 & -71.6100 &  2.27 &      & untr & 5.69  &    85 & 67    & GRB & GRB000416 / XRF \\
000424  & 9  &x& 9004&2 & 51658.762510 &       34.0 &     2.5  &  104.7273 & +49.8753 &  2.44 &      & 8086 & 3.39  &  n.a. &       & GRB & GRB000424 \\
000528  & 9  &+& 9145&2 & 51692.365500 &      108.5 &     5.9  &  161.2791 & -33.9859 &  1.60 &      &      & 2.65  &  1615 & 1.6   & GRB & GRB000528 \\
000529  & 9  &+& 9153&2 & 51693.363180 &       28.4 &     2.3  &    2.3505 & -61.5298 &  2.46 &      &      & 3.15  &   860 & 3.7   & GRB & GRB000529 \\
000608  & 9  &x& 9232&2 & 51703.368920 &       47.5 &    13.8  &    6.2623 & -69.0416 &  3.28 &      &      & 0.84  &    50 & 17    & GRB & GRB000608 \\
000615  & 9  &+& 9251&1 & 51710.262360 &       78.0 &     4.9  &  233.1700 & +73.8184 &  2.01 &      &      & 3.49  &   160 & 22    & GRB & GRB000615 / XRF \\
000619  & u  & & 9275&1 & 51714.359805 &    46040.4 &  2876.4  &   13.2643 & -74.6457 &  1.86 & 0.38 &      & 0.068 &       &       & RS CVn & CF Tuc \\
000620  & 9  &x& 9275&2 & 51715.231620 &       31.1 &     2.7  &  113.8024 & +69.1922 &       &      &      & 3.53  &   895 & 3.9   & GRB & GRB000620 \\
000627  & u  & & 9331&1 & 51722.501215 &     1656.5 &   657.9  &  341.7074 & +44.3234 &  2.33 & 1.17 &      & 0.087 &       &       & Flare star & EV Lac 2 / M-dwarf flare stars with superflares (Ref32) \\
000628  & u  & & 9341&2 & 51723.772880 &       80.0 &     9.7  &  236.1639 & -69.7477 &  2.82 &      &      & 0.84  &    50 & 17    & GRB & GRB000628 \\
000630  & u  & & 9352&1 & 51725.281238 &       72.3 &    16.3  &  233.8921 & +57.8930 &  2.82 &      &      & 1.65  &    75 & 22    & GRB & GRB000630 \\
000901  & u  & & 9696&2 & 51788.053540 &      520.3 &    20.7  &  268.6004 & -26.3313 &  4.4  &      &      & 0.91  & $<$20 & $>$46 & GRB? & FXT000901 / XRF, coincides with EQ J084834.9-785352 / M5 star (unknown as flare star). (Ref35) \\
000920  & 23 & & 9832&2 & 51807.316978 &    27644.8 &   347.7  &  268.6704 & -26.3206 &  1.87 & 3.42 &      & 0.263 &       &       & SFXT & IGR J17544-2619 flare 4 \\
001008  & 23 & & 9934&2 & 51825.121897 &     2845.9 &   720.2  &  268.6704 & -26.3206 &  2.72 & 0.48 &      & 0.272 &       &       & SFXT & IGR J17544-2619 flare 5 \\
001011  & 9  &x& 9957&2 & 51828.663080 &       48.2 &     2.2  &  275.7687 & -50.8999 &  1.56 & 0.53 &      & 3.60  &  3005 & 1.2   & GRB & GRB001011 \\
001024  & u  & &10050&1 & 51841.331705 &       17.6 &     0.8  &  125.5960 & -87.7158 &  3.80 &      &      & 1.96  &   105 & 19    & GRB & GRB001024 / XRF \\
001028  & u  & &10055&2 & 51845.510840 &       70.0 &   162.7  &  173.2039 & -86.1883 &  5.40 &      &      & 0.16  & $<$50 & $>$3.3& GRB? & FXT001028 / XRF \\
001101  & u  & &10091&1 & 51849.790690 &      120.3 &    17.6  &  123.1415 & +48.2890 &  2.27 &      &      & 0.74  & $<$10 & $>$74 & GRB? & FXT001101 \\
001109  & 9  &+&10146&1 & 51857.391080 &       74.9 &     2.9  &  277.5242 & +55.3079 &  1.84 &      &      & 2.86  &   460 & 6.2   & GRB & GRB001109 \\
001110  & u  & &10159&2 & 51858.592130 &       61.4 &    12.0  &  300.6197 & -20.5037 &  2.56 &      &      & 0.97  & $<$20 & $>$48 & GRB? & FXT001110 / XRF \\
001207  & u  & &10302&1 & 51885.179691 &     3213.1 &  1055.0  &   38.5894 & -43.7960 &  1.46 & 0.27 &      & 0.485 &       &       & Flare star & CC Eri, brightest stellar flare \\
001213  & u  & &10320&2 & 51891.353117 &    42078.5 &  2763.9  &  332.1754 & +45.7434 &  2.20 & 0.21 &      & 0.139 &       &       & RS CVn & AR Lac flare 1 \\
010119  & u  & &10570&1 & 51927.832386 &    63553.8 &  6590.2  &   47.0529 & +40.9608 &  1.83 & 0.58 &      & 0.115 &       &       & Flare star & Algol flare 3 / K-type flaring star in a binary with an early-type star \\
010213  & a  & &10725&2 & 51953.122980 &       22.6 &     2.4  &  257.3390 & +39.2681 &  1.93 &      &      & 5.80  &  1550 & 3.7   & GRB & GRB010213 / XRF \\
010214  & 36 &+&10727&2 & 51954.366670 &       78.8 &     4.3  &  265.2453 & +48.5508 &  2.25 &      &      & 1.98  &   655 & 3.0   & GRB & GRB010214 \\
010219  & u  & &10747&1 & 51959.258830 &       32.5 &     2.2  &   94.1820 & -69.7800 &  2.26 &      &      & 4.10  &   160 & 26    & GRB & GRB010219 \\
010220  & 9  &+&10753&2 & 51960.952140 &      247.0 &    21.4  &   39.2459 & +61.7612 &  1.97 &      &      & 0.66  &   455 & 1.5   & GRB & GRB010220 \\
010222  & 37 &+&10767&1 & 51962.306539 &      123.5 &     3.5  &  223.0499 & +43.0130 &  1.46 & 0.34 &      & 24.53 &  9840 & 2.5   & GRB & GRB010222 \\
010301  & u  & &10800&1 & 51969.647174 &     4924.1 &   383.8  &   82.1948 & -65.4533 &  1.68 & 0.28 &      & 0.267 &       &       & T Tau star & AB Dor flare 6 \\
010304  & 9  & &10822&2 & 51972.221380 &       18.5 &     1.3  &  316.5876 & +53.2077 &       &      &      & 4.99  &  1225 & 4.1   & GRB & GRB010304 \\
010304b & u  & &10822&1 & 51972.42479  &    40679.7 &  2875.1  &  130.2049 & -45.0477 &  2.76 & 0.69 &      & 0.101 &       &       & SFXT & IGR J08408-4503 \\
010315  & u  & &10883&1 & 51981.336717 &   178681.4 &  3009.2  &  175.0108 & -65.3811 &  1.55 & 3.55 &      & 0.064 &       &       & RS CVn & GT Mus \\
010320  & u  & &10923&1 & 51988.863100 &       71.0 &    13.7  &  277.1368 & -10.6302 &  2.84 &      &      & 1.12  & $<$30 & $>$37 & GRB? & FXT010320 \\
010324a & 38 & &10954&1 & 51992.480173 &      464.4 &    29.4  &  107.8541 & +19.9954 &  2.43 &      &      & 2.38  &   510 &   4.7 & GRB & GRB010324 / coincident with LEDA 1613994 \\
010324b & u  & &10954&2 & 51992.622915 &     3281.9 &   838.8  &  282.5536 &  -4.5534 &  9.70 & 2.04 &      & 0.068 &       &       & SFXT & AX J1845.0-0433 \\
010327  & u  & &10972&1 & 51995.627257 &    28185.2 &  2319.9  &   82.1861 & -65.4506 &  2.32 & 0.01 &      & 0.071 &       &       & T Tau star & AB Dor flare 7 \\
010412  & 9  &x&11085&1 & 52011.907160 &       88.8 &     3.1  &  294.9057 & +13.6124 &  1.65 &      &      & 4.74  &  1860 & 2.5   & GRB & GRB010412 \\
010501a & u  & &11203&2 & 52030.079540 &      215.2 &    19.2  &   88.2420 & +78.7952 &  2.31 &      &      & 0.30  &    60 & 4.9   & GRB & GRB010501a \\
010501b & 9  & &11203&1 & 52030.275950 &       71.9 &     6.8  &  286.7133 & -70.1803 &  4.18 &      &      & 0.34  &   120 & 2.9   & GRB & GRB010501b \\
010512  & u  & &11268&1 & 52041.02962  &   104431.7 &  1966.7  &  332.1789 & +45.7261 &  3.42 & 0.98 &      & 0.092 &       &       & RS CVn & AR Lac flare 2 \\
010518  & 9  &x&11292&2 & 52047.278780 &      264.2 &    23.6  &  161.6546 & -57.7643 &  3.21 &      &      & 0.43  &   130 & 3.3   & GRB & GRB010518 \\
010527  & u  & &11324&2 & 52056.347470 &       71.4 &    10.2  &  189.2384 & +36.1791 &  3.01 &      &      & 1.10  &    65 & 17    & GRB & GRB010527 \\
010528  & 39 & &11330&1 & 52057.734213 &     3100.8 &   690.1  &  116.9450 & -57.6173 &  2.41 & 0.51 &      & 0.178 &       &       & Flare star & V857 Cen = Gl 431 / M dwarf flare star \\
010707  & 9  & &11486&1 & 52097.287850 &       16.4 &     2.7  &  205.5666 & -38.5365 &  2.16 &      &      & 2.04  & $<$70 & $>$29 & GRB? & FXT010707 / XRF \\
010824  & u  & &11828&2 & 52145.123067 &     1566.4 &   530.7  &   56.6818 & -39.1525 & 14.19 &      &      & 0.243 &       &       & T Tauri? & SAX J0346-3906 / see also 990810 \\
011030  & 9  &x&12150&1 & 52212.268328 &      280.3 &    14.2  &  310.9180 & +77.2925 &  1.95 & 0.50 &      & 0.52  &    30 & 18    & GRB & GRB011030 / XRR, first real time follow up X-ray flash \\
011103  & u  & &12167&2 & 52216.139073 &    23581.1 &  3016.2  &  332.1717 & +45.6598 &  2.86 & 4.90 &      & 0.067 &       &       & RS CVn & AR Lac flare 3 \\
011121  & 9  &+&12252&1 & 52234.782435 &      284.3 &     2.4  &  173.6086 & -76.0259 &  1.45 & 0.26 &      & 29.64 &  7695 & 3.9   & GRB & GRB011121 / 2nd brightest IN WFC / peak flux of first peak\\
011211  & 9  &+&12361&1 & 52254.795040 &      276.2 &    18.2  &  168.8179 & -21.9368 &  1.70 & 0.83 &      & 0.72  &    50 & 14    & GRB & GRB011211 / XRR \\
020113  & u  & &12497&2 & 52287.887420 &      909.4 &    62.3  &  226.6594 & +83.4713 &  4.12 &      &      & 0.18  &    35 & 5.2   & GRB & GRB020113 \\
020118  & 9  & &12513&2 & 52292.769646 &       54.3 &     4.4  &    9.4460 & +59.5364 &  2.27 &      &      & 2.85  &    60 & 47    & GRB & GRB020118 \\
020321  & 31 &+&12705&1 & 52354.180980 &      135.9 &    13.4  &  243.1410 & -83.7157 &  3.31 &      &      & 0.31  &   100 & 3.1   & GRB & GRB020321 \\
020322  & 3  &+&12712&2 & 52355.160819 &      105.7 &     6.0  &  270.2379 & +81.0880 &  2.47 & 1.21 &      & 0.75  &    80 & 9.4   & GRB & GRB020322 \\
020409  & 3  &x&12830&1 & 52373.882928 &       73.5 &     4.9  &  131.2984 & +66.6810 &  3.75 &      &      & 3.24  &   210 & 15    & GRB & GRB020409 \\
020410  &40  &+&12830&2 & 52374.444190 &     1137.8 &    10.8  &  331.6779 & -83.8217 &  1.52 & 0.34 &      & 1.46  &  n.a. &       & GRB? & GRB020410 / XRR \\
020427  &41,42&+&12933&2& 52391.158859 &       50.4 &     4.6  &  332.3803 & -65.3276 &  2.22 & 0.34 &      & 1.90 & $<$100 & $>$19 & GRB? & GRB020427 / XRF, first coincidence of GRB with ionospheric event \\
\hline
\end{longtable}
\noindent
\raggedright $^\dag$1 - \cite{piro1998};
2 - \cite{zand2004}; 
3 - \cite{zand2004b};
4 - \cite{costa1997b};
5 - \cite{costa1997};
6 - \cite{nicastro1998};
7 - \cite{piro1998};
8 - \cite{kippen2003};
9 - \cite{vetere2007};
10 - \cite{dalfiume2000};
11 - \cite{antonelli1999};
12 - \cite{zand1998};
13 - \cite{zand1998b}
14 - \cite{celidonio1998};
15 - \cite{zand1998};
16 - \cite{pian1999};
17 - \cite{romano2008};
18 - \cite{ricci1998};
19 - \cite{zand1998};
20 - \cite{frontera2000};
21 - \cite{corsi2005};
22 - \cite{zand2000b};
23 - \cite{zand2004};
24 - \cite{kuulkers2000};
25 - \cite{piro1999};
26 - \cite{gandolfi1999};
27 - \cite{feroci2001};
28 - \cite{amati2000};
29 - \cite{frontera2001};
30 - \cite{montanari2001};
31 - \cite{zand2000};
32 - \cite{muller1999};
33 - \cite{heise2001};
34 - \cite{piro2002};
35 - \cite{antonelli2000}; 
36 - \cite{guidorzi2003};
37 - \cite{zand2001};
38 - \cite{guidorzi2001};
39 - \cite{zand2001};
40 - \cite{nicastro2004};
41 - \cite{amati2004};
42 - \cite{fishman2002};
43 - \cite{osten2010};
44 - \cite{oord1989};
45 - \cite{seli2025};
46 - \cite{lawson2002}.
\end{landscape}
}

\footnotesize
\longtab[2]{
\begin{longtable}{lrclrrlllrll}
\caption{Spectral fits with a power law. No line was included in fit,
  except for events listed in Table~\ref{table3}. $N_{\rm H}$ was
  fixed at the Galactic values, except for events listed in
  Table~\ref{table4}. Distances were extracted from Simbad and are in
  parsec via GAIA parallaxes or redshift (indicated with prefix
  $z=$).}
\label{table2}
\tabularnewline
\hline\hline
ID & \multicolumn{1}{l}{OP}&WFC & MJD-start & Time & $N_{\rm H}$ &Photon & 2-30 keV           & $\chi^2_{\rm r}$   & Distance & 2-30 keV & Identification \\
   &   &    &           & span & (10$^{20}$  & index & fluence ($10^{-6}$ &                  & (pc/z)    & energy   &               \\
   &   &    &           & (s)  & cm$^{-2}$)    &       & erg cm$^{-2}$)     &                  &            & (ergs)  &   \\
\hline
\endfirsthead
\caption{continued.}\\
ID & \multicolumn{1}{l}{OP}&WFC & MJD-start & Time & $N_{\rm H}$ &Photon & 2-30 keV           & $\chi^2_{\rm r}$   & Distance & 2-30 keV & Identification \\
   &   &    &           & span & (10$^{21}$  & index & fluence ($10^{-6}$ &                  & (pc/z)    & energy   &               \\
   &   &    &           & (s)  & cm$^{-2}$)    &       & erg cm$^{-2}$)     &                  &            & (ergs)  &   \\
\hline\hline
\endhead
\hline
\endfoot
960707  &   616&1 & 50271.419483 & 5400   &   1.50 & 2.49$\pm$0.09 &  2.96$\pm$0.14 & 1.38 &  5.05    & 1.5$\times10^{34}$ & EV Lac 1 \\
960720  &   675&1 & 50284.483892 & 30     &   0.24 & 1.89$\pm$0.20 &  0.33$\pm$0.05 & 0.94 &          &                    & GRB960720 \\
960726  &   689&2 & 50290.495360 & 100    &   2.06 & 2.10$\pm$0.18 &  1.16$\pm$0.14 & 0.74 &          &                    & GRB960726 \\
960806  &   735&1 & 50301.266350 & 40     &   0.25 & 1.81$\pm$0.19 &  0.33$\pm$0.05 & 0.62 &          &                    & GRB960806 \\
960813  &   757&2 & 50308.952389 & 3000   &   5.30 & 2.24$\pm$0.26 &  1.69$\pm$0.29 & 0.76 &          &                    & EE Mus \\
960825  &   820&2 & 50320.514881 & 14400  &  13.20 & 1.71$\pm$0.14 & 23.59$\pm$2.05 & 0.89 & 2523.34  & 2.7$\times10^{40}$ & IGR J17544-2619 1 \\
961005  &  1075&2 & 50361.247077 & 18000  &   1.00 & 1.99$\pm$0.08 & 19.94$\pm$1.17 & 1.60 &  27.57   & 2.5$\times10^{36}$ & Algol 1 \\
961229a &  1448&2 & 50446.555400 & 80     &   0.29 & 0.79$\pm$0.37 &  0.36$\pm$0.16 & 1.00 &          &                    & FXT961229a \\
961229b &  1448&1 & 50446.726181 & 380    &   0.25 & 1.70$\pm$0.25 &  1.74$\pm$0.36 & 0.60 &          &                    & FXT961229b \\
970111a &  1485&2 & 50459.405560 & 70     &   0.45 & 0.65$\pm$0.07 &  7.17$\pm$0.44 & 0.71 &          &                    & GRB970111 \\
970111b &  1487&1 & 50459.983632 & 2800   &   0.36 & 1.50$\pm$0.06 & 10.92$\pm$0.50 & 1.09 &          &                    & GRB970111b \\
970123  &  1565&2 & 50471.857736 & 5400   &   0.50 & 2.22$\pm$0.11 &  4.27$\pm$0.26 & 1.40 &  16.56   & 1.7$\times10^{35}$ & BY Dra 1 \\
970228  &  1671&1 & 50507.123620 & 75     &   1.36 & 1.62$\pm$0.08 &  3.52$\pm$0.24 & 0.85 &$z$=0.695 & 2.3$\times10^{51}$ & GRB970228 \\
970321  &  1780&1 & 50528.194202 & 5400   &   0.80 & 2.44$\pm$0.13 &  3.45$\pm$0.21 & 0.98 &  15.55   & 1.1$\times10^{35}$ & NLTT 51688 \\
970402$^\dag$&1830&1&50540.929040 & 130    &   1.69 & 1.07$\pm$0.10 &  1.62$\pm$0.14 & 0.67 &          &                    & GRB970402 \\
970427a &  1930&2 & 50565.015468 & 72000  &   3.60 & 1.99$\pm$0.19 &  6.75$\pm$0.94 & 0.63 &  412.39  & 2.1$\times10^{38}$ & V969 Mon \\
970508  &  1984&2 & 50576.904010 & 20     &   0.40 & 1.75$\pm$0.09 &  0.66$\pm$0.05 & 0.56 &$z$=0.835 & 9.9$\times10^{50}$ & GRB970508 \\
971010  &  2622&2 & 50731.675460 & 3000   &   0.40 & 2.36$\pm$0.37 &  1.64$\pm$0.41 & 1.20 &  10.70   & 2.4$\times10^{34}$ & AT Mic \\
971018  &  2699&2 & 50739.377130 & 10800  &   0.70 & 2.26$\pm$0.13 &  2.98$\pm$0.23 & 1.04 &  14.85   & 1.3$\times10^{35}$ & AB Dor 1 \\
971019a &  2705&2 & 50740.664495 & 23     &   0.21 & 1.21$\pm$0.15 &  1.54$\pm$0.22 & 1.09 &          &                    & XRF971019a \\
971019b &  2706&1 & 50740.738210 & 200    &   0.92 & 1.78$\pm$0.28 &  0.39$\pm$0.09 & 0.84 &          &                    & GRB971019b \\
971024  &  2732&1 & 50745.783860 & 35     &   0.14 & 1.69$\pm$0.12 &  0.41$\pm$0.04 & 0.88 &          &                    & GRB971024 \\
971206a &  3066&2 & 50788.206717 & 108000 &   0.80 & 2.07$\pm$0.14 & 13.62$\pm$1.28 & 1.16 &  98.37   & 3.4$\times10^{37}$ & IM Peg \\
971206b &  3066&1 & 50788.805960 & 60     &   0.39 & 1.35$\pm$0.27 &  0.79$\pm$0.20 & 0.58 &          &                    & GRB971226 \\
971214  &  3123&1 & 50796.972640 & 80     &   0.14 & 0.62$\pm$0.25 &  1.20$\pm$0.30 & 0.48 &$z$=3.418 & 3.0$\times10^{52}$ & GRB971214 \\
971227  &  3205&2 & 50809.349380 & 6      &   0.09 & 0.46$\pm$0.19 &  0.39$\pm$0.07 & 0.62 &          &                    & GRB971227 \\
980109  &  3289&2 & 50822.050240 & 50     &   0.14 & 1.11$\pm$0.12 &  1.11$\pm$0.12 & 0.91 &          &                    & GRB980109 \\
980128  &  3465&2 & 50841.039070 & 130    &   3.61 & 1.28$\pm$0.15 &  0.49$\pm$0.06 & 0.78 &          &                    & GRB980128 \\
980218  &  3714&1 & 50862.752755 & 7200   &   0.70 & 1.97$\pm$0.19 &  4.17$\pm$0.56 & 0.45 &  14.85   & 1.5$\times10^{35}$ & AB Dor 2 \\
980306  &  3875&2 & 50878.161680 & 70     &   0.70 & 2.15$\pm$0.21 &  0.18$\pm$0.03 & 0.48 &          &                    & GRB980306 \\
980311  &  3905&2 & 50883.855408 & 10800  &  11.70 & 2.28$\pm$0.73 &  8.83$\pm$8.83 & 1.53 &  2469.14 & 1.1$\times10^{40}$ & SAX J1818.6-1703 \\
980326  &  4057&1 & 50898.888130 & 8      &   0.63 & 1.07$\pm$0.20 &  0.69$\pm$0.13 & 1.23 &$z$=0.9   & 1.2$\times10^{51}$ & GRB980326 \\
980329  &  4073&2 & 50901.155986 & 30     &   0.88 & 0.70$\pm$0.13 &  4.34$\pm$0.54 & 1.12 &$z$=3.6   & 1.2$\times10^{53}$ & GRB980329 \\
980412  &  4169&2 & 50915.295220 & 300    &   2.28 & 1.95$\pm$0.42 &  0.34$\pm$0.12 & 0.56 &  408.11  & 6.8$\times10^{36}$ & FXT980412 \\
980415  &  4202&2 & 50918.645360 & 100    &   0.23 & 1.08$\pm$0.37 &  0.30$\pm$0.13 & 1.20 &          &                    & FXT980415 \\
980415b &  4202&2 & 50918.711636 & 50400  &   0.30 & 2.41$\pm$0.09 &  9.63$\pm$0.49 & 1.73 &  138.89  & 3.2$\times10^{37}$ & GR Dra 1 \\
980425  &  4281&2 & 50928.909146 & 50     &   0.38 & 1.53$\pm$0.17 &  2.30$\pm$0.32 & 0.82 &$z$=0.0085& 3.6$\times10^{47}$ & GRB980425 \\
980429  &  4320&2 & 50932.365567 & 180    &   0.30 & 1.84$\pm$0.09 &  2.65$\pm$0.19 & 1.39 &          &                    & GRB980429 \\
980515  &  4462&2 & 50948.708330 & 50     &   0.28 & 0.98$\pm$0.16 &  1.31$\pm$0.18 & 0.94 &          &                    & GRB980515 \\
980519  &  4480&2 & 50952.512950 & 270    &   1.85 & 1.67$\pm$0.03 &  4.85$\pm$0.14 & 0.67 &$z$=1.58  & 2.6$\times10^{52}$ & GRB980519 \\
980613  &  4677&2 & 50977.202150 & 40     &   0.47 & 1.58$\pm$0.24 &  0.29$\pm$0.06 & 1.06 &$z$=1.0969& 7.4$\times10^{50}$ & GRB980613 \\
980614  &  4683&1 & 50978.359200 & 350    &   0.72 & 0.95$\pm$0.40 &  1.15$\pm$0.58 & 0.71 &          &                    & FXT980614 \\
980706  &  4894&2 & 51000.410380 & 50     &   0.10 & 1.84$\pm$0.23 &  0.24$\pm$0.04 & 0.82 &          &                    & FXT980706 \\
980804  &  5072&2 & 51029.678414 & 50400  &   1.20 & 1.80$\pm$0.08 & 14.15$\pm$0.82 & 1.23 &  50.55   & 6.1$\times10^{36}$ & UX Ari \\
980824  &  5166&2 & 51049.607200 & 60     &   0.30 & 2.44$\pm$0.39 &  0.13$\pm$0.03 & 1.05 &          &                    & FXT980824 \\
980907  &  5275&1 & 51063.014660 & 1200   &   1.40 & 1.39$\pm$0.22 &  2.03$\pm$0.37 & 0.58 &  10.89   & 3.3$\times10^{34}$ & NLTT18592 \\
980924  &  5385&1 & 51080.359688 & 3000   &   0.50 & 2.03$\pm$0.19 &  2.03$\pm$0.28 & 1.07 &  16.56   & 6.7$\times10^{34}$ & BY Dra 2 \\
981018  &  5618&2 & 51104.018940 & 420    &   1.50 & 1.61$\pm$0.14 &  3.29$\pm$0.37 & 0.80 &          &                    & GRB981018 \\
981102  &  5745&1 & 51119.770233 & 28800  &   0.70 & 2.21$\pm$0.17 &  3.91$\pm$0.42 & 1.08 &  14.85   & 2.1$\times10^{35}$ & AB Dor 3 \\
981217  &  6002&2 & 51164.079960 & 160    &   0.56 & 1.21$\pm$0.09 &  1.95$\pm$0.15 & 0.82 &          &                    & FXT981217 \\
981226  &  6055&1 & 51173.407872 & 250    &   0.15 & 1.86$\pm$0.09 &  1.14$\pm$0.07 & 0.44 &$z$=1.11  & 2.2$\times10^{51}$ & GRB981226 \\
981227  &  6069&2 & 51174.430038 & 3000   &  12.10 & 1.83$\pm$0.12 &  2.44$\pm$0.20 & 1.19 &  109.27  & 5.0$\times10^{36}$ & V1373 Cen \\
990123  &  6221&1 & 51201.407801 & 80     &   0.18 & 0.86$\pm$0.02 & 12.92$\pm$0.24 & 1.38 &$z$=1.6   & 7.1$\times10^{52}$ & GRB990123 \\
990125  &  6231&2 & 51203.466428 & 21600  &   1.00 & 2.31$\pm$0.07 &  6.70$\pm$0.28 & 0.60 &  27.57   & 1.3$\times10^{36}$ & Algol 2 \\
990217  &  6383&2 & 51226.224560 & 100    &   0.12 & 1.79$\pm$0.14 &  0.44$\pm$0.04 & 0.86 &          &                    & GRB990217 \\
990220a &  6419&1 & 51229.584397 & 10800  &   0.30 & 5.09$\pm$1.40 &  1.17$\pm$1.17 & 1.80 &  5910.17 & 9.6$\times10^{39}$ & SAX J1819.3-2525 3 \\
990220b &  6419&1 & 51229.728314 & 720    &  13.20 & 0.97$\pm$0.11 &  4.43$\pm$0.35 & 0.93 &  2523.34 & 3.4$\times10^{39}$ & IGR J17544-2619 2 \\
990227  &  6473&2 & 51236.778419 & 18000  &   0.30 & 1.84$\pm$0.12 &  9.90$\pm$0.86 & 0.77 &  138.89  & 3.0$\times10^{37}$ & GR Dra 2 \\
990311  &  6570&1 & 51248.754346 & 2400   &  13.20 & 1.99$\pm$0.24 &  2.74$\pm$0.51 & 0.41 &  2523.34 & 2.1$\times10^{39}$ & IGR J17544-2619 6 \\
990328  &  6742&1 & 51265.715390 & 250    &   3.57 & 1.82$\pm$0.29 &  0.32$\pm$0.07 & 0.94 &          &                    & FXT990328 \\
990413  &  6829&2 & 51281.232400 & 840    &   0.23 & 1.01$\pm$0.15 &  1.55$\pm$0.20 & 0.42 &          &                    & GRB990413 \\
990510  &  6902&2 & 51308.367450 & 120    &   0.82 & 1.36$\pm$0.06 &  6.55$\pm$0.32 & 0.93 &$z$=1.6187& 3.7$\times10^{52}$ & GRB990510 \\
990520  &  6952&1 & 51318.085320 & 15     &   0.37 & 1.56$\pm$0.17 &  0.31$\pm$0.04 & 0.54 &          &                    & GRB990520 \\
990526  &  6975&1 & 51324.515180 & 40     &   3.39 & 2.61$\pm$0.33 &  0.19$\pm$0.04 & 1.28 &          &                    & GRB990526 \\
990531  &  6990&1 & 51329.665601 & 1800   &   0.60 & 1.87$\pm$0.16 &  4.10$\pm$0.51 & 0.88 &  49.67   & 1.4$\times10^{36}$ & CD-33 7795 \\
990625  &  7072&2 & 51354.016980 & 20     &   0.13 & 1.69$\pm$0.17 &  0.59$\pm$0.08 & 0.97 &          &                    & GRB990625 \\
990626  &  7088&2 & 51355.664400 & 2400   &   0.40 & 1.88$\pm$0.11 &  5.36$\pm$0.42 & 0.93 &  6.18    & 2.5$\times10^{34}$ & EQ Peg \\
990627  &  7096&2 & 51356.208940 & 40     &   0.60 & 1.15$\pm$0.12 &  0.44$\pm$0.04 & 0.78 &          &                    & GRB990627 \\
990704  &  7141&1 & 51363.729230 & 30     &   0.28 & 1.58$\pm$0.06 &  3.17$\pm$0.15 & 1.29 &          &                    & GRB990704 \\
990705  &  7147&2 & 51364.667650 & 50     &   1.04 & 0.88$\pm$0.07 &  7.97$\pm$0.60 & 1.46 &$z$=0.843 & 1.2$\times10^{52}$ & GRB990705 \\
990709  &  7167&1 & 51368.803083 & 11880  &   0.50 & 1.90$\pm$0.21 &  3.51$\pm$0.55 & 0.91 &  39.12   & 1.2$\times10^{36}$ & II Peg \\
990712  &  7188&2 & 51371.696550 & 40     &   0.26 & 1.79$\pm$0.04 &  5.10$\pm$0.17 & 1.46 &$z$=0.433 & 2.0$\times10^{51}$ & GRB990712 \\
990722  &  7263&2 & 51381.388974 & 50400  &   0.50 & 4.76$\pm$2.39 &  0.47$\pm$0.47 & 0.92 &  39.12   & 1.2$\times10^{35}$ & II Peg \\
990806  &  7348&1 & 51396.602860 & 35     &   0.49 & 1.44$\pm$0.12 &  0.85$\pm$0.08 & 0.66 &          &                    & GRB990806 \\
990810  &  7362&2 & 51400.612230 & 1200   &   0.30 & 1.65$\pm$0.13 &  2.37$\pm$0.25 & 1.44 &  44.01   & 5.8$\times10^{35}$ & SAX J0346.6-3906 \\
990907  &  7523&2 & 51428.732694 & 190    &   1.35 & 1.03$\pm$0.10 &  2.09$\pm$0.18 & 1.05 &          &                    & GRB990907 \\
990908  &  7523&2 & 51429.012500 & 130    &   0.98 & 1.63$\pm$0.04 &  2.60$\pm$0.08 & 1.25 &          &                    & GRB990908 \\
990909  &  7546&2 & 51430.894314 & 28800  &   0.20 & 1.65$\pm$0.03 & 85.46$\pm$1.98 & 1.24 &  5910.17 & 5.7$\times10^{41}$ & SAX J1819.3-2525 2 \\
990911  &  7553&1 & 51432.115942 & 28800  &  13.20 & 1.24$\pm$0.09 & 24.20$\pm$1.52 & 0.67 &  2523.34 & 3.8$\times10^{40}$ & IGR J17544-2619 3 \\
991005  &  7692&2 & 51456.046632 & 9000   &   0.70 & 2.10$\pm$0.14 &  4.91$\pm$0.45 & 0.98 &  14.85   & 2.0$\times10^{35}$ & AB Dor 4 \\
991014  &  7749&2 & 51465.911500 & 7      &   2.12 & 0.55$\pm$0.23 &  0.42$\pm$0.10 & 0.63 &          &                    & GRB991014 \\
991026  &  7820&2 & 51477.543218 & 95     &   0.94 & 1.44$\pm$0.06 &  2.72$\pm$0.14 & 1.29 &          &                    & GRB991026 \\
991030  &  7840&2 & 51481.073224 & 120    &   2.52 & 1.18$\pm$0.13 &  1.80$\pm$0.21 & 0.65 &          &                    & GRB991030 \\
991105  &  7859&2 & 51487.694930 & 100    &   3.88 & 1.40$\pm$0.13 &  1.39$\pm$0.16 & 0.90 &          &                    & GRB991105 \\
991106  &  7860&1 & 51488.454310 & 25     &   4.43 & 1.60$\pm$0.21 &  0.19$\pm$0.03 & 0.85 &          &                    & GRB991106 \\
991217  &  8073&2 & 51529.174610 & 100    &   0.39 & 1.59$\pm$0.10 &  0.70$\pm$0.06 & 1.21 &          &                    & FXT991217 \\
000206  &  8374&1 & 51580.866460 & 40     &   0.21 & 1.72$\pm$0.09 &  0.59$\pm$0.04 & 0.54 &          &                    & GRB000206 \\
000208  &  8386&1 & 51582.181430 & 60     &   0.12 & 1.77$\pm$0.22 &  0.25$\pm$0.04 & 0.56 &          &                    & GRB000208 \\
000210  &  8406&1 & 51584.363870 & 30     &   0.18 & 0.91$\pm$0.07 &  4.91$\pm$0.31 & 0.79 &  $z$=0.8456& 7.5$\times10^{51}$ & GRB000210 \\
000214  &  8425&1 & 51588.042360 & 100    &   0.45 & 1.88$\pm$0.07 &  1.78$\pm$0.09 & 0.99 &  $z$=0.46  & 8.1$\times10^{50}$  & GRB000214 \\
000218  &  8468&2 & 51592.667380 & 505    &   1.07 & 1.72$\pm$0.11 &  3.51$\pm$0.30 & 0.55 &          &                    & GRB000218 \\
000312  &  8683&2 & 51615.495490 & 2400   &   0.40 & 2.21$\pm$0.17 &  1.73$\pm$0.19 & 0.87 &  11.70   & 3.1$\times10^{34}$ & CD-35 2213 \\
000409  &  8881&1 & 51643.237951 & 3600   &   0.70 & 2.01$\pm$0.12 &  2.31$\pm$0.18 & 1.03 &  14.85   & 1.0$\times10^{35}$ & AB Dor 5 \\
000416  &  8942&2 & 51650.606060 & 50     &   0.53 & 2.22$\pm$0.12 &  1.02$\pm$0.08 & 0.84 &          &                    & GRB000416 \\
000424  &  9004&2 & 51658.762510 & 20     &   0.83 & 1.04$\pm$0.14 &  0.80$\pm$0.10 & 0.87 &          &                    & GRB000424 \\
000528  &  9145&2 & 51692.365500 & 110    &   0.72 & 0.56$\pm$0.04 &  4.54$\pm$0.17 & 2.32 &          &                    & GRB000528 \\
000529  &  9153&2 & 51693.363410 & 30     &   0.14 & 1.46$\pm$0.24 &  0.31$\pm$0.06 & 0.98 &          &                    & GRB000529 \\
000608  &  9232&2 & 51703.368920 & 50     &   0.20 & 2.36$\pm$0.25 &  0.32$\pm$0.05 & 0.81 &          &                    & GRB000608 \\
000615  &  9251&1 & 51710.262360 & 100    &   0.23 & 1.85$\pm$0.12 &  2.24$\pm$0.20 & 0.89 &          &                    & GRB000615 \\
000619  &  9275&1 & 51714.359805 & 54000  &   0.90 & 1.77$\pm$0.08 & 15.14$\pm$1.17 & 1.43 &  87.53   & 2.1$\times10^{37}$ & CF Tuc \\
000620  &  9275&2 & 51715.231620 & 21     &   0.40 & 1.02$\pm$0.26 &  0.95$\pm$0.24 & 1.36 &          &                    & GRB000620 \\
000627  &  9331&1 & 51722.501215 & 2400   &   1.50 & 2.29$\pm$0.15 &  1.58$\pm$0.14 & 1.48 &  5.05    & 5.1$\times10^{33}$ & EV Lac 2 \\
000628  &  9341&2 & 51723.772880 & 60     &   0.82 & 1.45$\pm$0.19 &  0.75$\pm$0.12 & 0.56 &          &                    & GRB000628 \\
000630  &  9352&1 & 51725.281240 & 70     &   0.22 & 1.85$\pm$0.17 &  0.82$\pm$0.11 & 0.93 &          &                    & GRB000630 \\
000901$^\ddag$&9696&2&51788.053540& 70     &  13.30 & 1.01$\pm$0.62 &  0.34$\pm$0.34 & 1.20 &  28.77   & 3.4$\times10^{34}$ & FXT000901 \\
000920  &  9832&2 & 51807.316978 & 36000  &  13.20 & 4.15$\pm$0.71 & 10.80$\pm$10.8 & 1.69 &  2523.34 & 1.2$\times10^{40}$ & IGR J17544-2619 4 \\
001008  &  9934&2 & 51825.121897 & 2000   &  13.20 & 1.48$\pm$0.15 &  4.39$\pm$0.52 & 1.38 &  2523.34 & 3.4$\times10^{39}$ & IGR J17544-2619 5 \\
001011  &  9957&2 & 51828.663070 & 60     &   0.67 & 1.28$\pm$0.05 &  2.02$\pm$0.08 & 0.90 &          &                    & GRB001011 \\
001024  & 10050&1 & 51841.331705 & 10     &   0.91 & 1.30$\pm$0.28 &  0.16$\pm$0.04 & 1.22 &          &                    & GRB001024 \\
001028  & 10055&2 & 51845.510840 & 120    &   0.87 & 0.94$\pm$0.48 &  0.23$\pm$0.17 & 0.79 &          &                    & FXT001028 \\
001101  & 10091&1 & 51849.790690 & 150    &   0.47 & 2.39$\pm$0.16 &  0.59$\pm$0.05 & 0.90 &          &                    & FXT001101 \\
001109  & 10146&1 & 51857.391080 & 65     &   0.34 & 1.22$\pm$0.08 &  1.90$\pm$0.14 & 0.94 &          &                    & GRB001109 \\
001110  & 10159&2 & 51858.592130 & 60     &   0.69 & 1.40$\pm$0.16 &  0.84$\pm$0.12 & 0.62 &          &                    & FXT001110 \\
001207  & 10302&1 & 51885.179691 & 2400   &   0.20 & 2.20$\pm$0.04 &  6.42$\pm$0.15 & 1.88 &  11.55   & 1.4$\times10^{35}$ & CC Eri \\
001213  & 10320&2 & 51891.353117 & 12600  &   2.30 & 1.71$\pm$0.13 &  6.60$\pm$0.69 & 1.30 &  42.51   & 2.9$\times10^{36}$ & AR Lac 1 \\
010119  & 10570&1 & 51927.832386 & 86400  &   1.00 & 2.34$\pm$0.11 &  9.43$\pm$0.55 & 0.94 &  27.57   & 2.1$\times10^{36}$ & Algol 3 \\
010213  & 10725&2 & 51953.123154 & 25     &   0.35 & 1.01$\pm$0.09 &  2.00$\pm$0.18 & 0.90 &          &                    & GRB010213 \\
010214  & 10727&2 & 51954.366670 & 40     &   0.20 & 0.95$\pm$0.11 &  0.88$\pm$0.08 & 0.67 &          &                    & GRB010214 \\
010219  & 10747&1 & 51959.258830 & 30     &   0.70 & 2.19$\pm$0.17 &  0.90$\pm$0.10 & 0.80 &          &                    & GRB010219 \\
010220  & 10753&2 & 51960.952140 & 200    &   7.08 & 0.83$\pm$0.10 &  1.78$\pm$0.15 & 1.41 &          &                    & GRB010220 \\
010222  & 10767&1 & 51962.306539 & 200    &   0.18 & 1.49$\pm$0.03 & 24.30$\pm$0.52 & 2.11 &$z$=1.477 & 1.1$\times10^{53}$ & GRB010222 \\
010301  & 10800&1 & 51969.647174 & 2400   &   0.70 & 2.02$\pm$0.07 &  5.10$\pm$0.25 & 1.65 &  14.85   & 1.5$\times10^{35}$ & AB Dor 6 \\
010304  & 10822&2 & 51972.221390 & 20     &   8.77 & 0.75$\pm$0.09 &  1.68$\pm$0.13 & 0.97 &          &                    & GRB010304 \\
010304b & 10822&1 & 51972.42479  & 50400  &   9.70 & 1.70$\pm$0.29 &  8.12$\pm$2.02 & 1.84 &          &                    & IGR J08408-4503 \\
010315$^\S$  & 10883&1 & 51981.336717 & 234000 &   7.50 & 1.91$\pm$0.03 & 47.95$\pm$1.02 &2.42 &119.09& 2.1$\times10^{38}$ & GT Mus \\
010320  & 10923&1 & 51988.863100 & 31     &  18.40 & 1.25$\pm$0.21 &  0.69$\pm$0.13 & 1.39 &          &                    & FXT010320 \\
010324a & 10954&1 & 51992.480173 & 400    &   0.98 & 1.43$\pm$0.15 &  7.78$\pm$0.99 & 1.36 &          &                    & GRB010324 \\
010324b & 10954&2 & 51992.622915 & 3300   &  14.00 & 2.46$\pm$0.41 &  1.97$\pm$1.37 & 1.47 &  6097.56 & 9.5$\times10^{39}$ & AX J1845.0-0433 \\
010327  & 10972&1 & 51995.627257 & 10800  &   0.70 & 2.32$\pm$0.18 &  2.63$\pm$0.26 & 0.80 &  14.85   & 1.1$\times10^{35}$ & AB Dor 7 \\
010412  & 11085&1 & 52011.907276 & 100    &   3.13 & 1.18$\pm$0.06 &  4.87$\pm$0.26 & 0.81 &          &                    & GRB010412 \\
010501a & 11203&2 & 52030.080697 & 200    &   0.62 & 1.19$\pm$0.14 &  0.64$\pm$0.07 & 1.36 &          &                    & GRB010501a \\
010501b & 11203&1 & 52030.276181 & 40     &   0.42 & 0.19$\pm$0.25 &  0.51$\pm$0.14 & 0.63 &          &                    & GRB010501b \\
010512  & 11268&1 & 52041.029624 & 3600   &   2.30 & 2.29$\pm$0.30 &  1.74$\pm$0.35 & 1.40 &  42.51   & 5.1$\times10^{35}$ & AR Lac 2 \\
010518  & 11292&2 & 52047.278780 & 300    &   7.29 & 1.30$\pm$0.24 &  0.92$\pm$0.19 & 1.20 &          &                    & GRB010518 \\
010527  & 11324&2 & 52056.347470 & 90     &   0.13 & 1.76$\pm$0.21 &  0.96$\pm$0.16 & 1.26 &          &                    & GRB010527 \\
010528  & 11330&1 & 52057.734213 & 2100   &   0.80 & 2.15$\pm$0.16 &  2.35$\pm$0.27 & 0.66 &  10.76   & 3.5$\times10^{34}$ & V857 Cen \\
010707  & 11486&1 & 52097.287850 & 20     &   0.41 & 1.83$\pm$0.12 &  0.30$\pm$0.03 & 2.07 &          &                    & FXT010707 \\
010824  & 11828&2 & 52145.123067 & 1500   &   0.20 & 1.23$\pm$2.89 &  0.11$\pm$0.11 & 0.99 &  44.01   & 3.0$\times10^{34}$ & SAX J0346-3906 \\
011030  & 12150&1 & 52212.268328 & 350    &   0.90 & 1.72$\pm$0.10 &  1.10$\pm$0.08 & 1.01 &          &                    & GRB011030 \\
011103  & 12167&2 & 52216.139073 & 5400   &   2.30 & 2.17$\pm$0.24 &  2.69$\pm$0.44 & 0.92 &  42.51   & 7.9$\times10^{35}$ & AR Lac 3 \\
011121  & 12252&1 & 52234.782435 & 100    &   0.78 & 1.37$\pm$0.03 & 14.62$\pm$0.37 & 0.94 &  $z$=0.36& 4.1$\times10^{51}$ & GRB011121 \\
011211  & 12361&1 & 52254.795040 & 400    &   0.34 & 1.43$\pm$0.07 &  2.69$\pm$0.16 & 0.86 &  $z$=2.1435& 2.6$\times10^{52}$ & GRB011211 \\
020113  & 12497&2 & 52287.887420 & 800    &   0.45 & 1.59$\pm$0.20 &  0.63$\pm$0.10 & 1.67 &          &                    & GRB020113 \\
020118  & 12513&2 & 52292.769646 & 60     &   3.91 & 1.39$\pm$0.15 &  1.18$\pm$0.14 & 0.60 &          &                    & GRB020118 \\
020321  & 12705&1 & 52354.180980 & 60     &   0.73 & 1.00$\pm$0.24 &  0.31$\pm$0.07 & 0.91 &          &                    & GRB020321 \\
020322  & 12712&2 & 52355.160819 & 60     &   0.40 & 1.40$\pm$0.14 &  0.65$\pm$0.08 & 0.80 &  $z$=1.80& 4.5$\times10^{51}$ & GRB020322 \\
020409  & 12830&1 & 52373.882928 & 50     &   0.33 & 1.26$\pm$0.23 &  1.00$\pm$0.22 & 0.56 &          &                    & GRB020409 \\
020410  & 12830&2 & 52374.444190 & 1300   &   0.70 & 1.81$\pm$0.05 &  9.59$\pm$0.34 & 1.00 &          &                    & GRB020410 \\
020427  & 12933&2 & 52391.158813 & 70     &   0.24 & 2.11$\pm$0.14 &  0.62$\pm$0.06 & 0.43 &  z$<$2.30&$<$7.0$\times10^{51}$& GRB020427 \\
\hline
\end{longtable}

\noindent
\raggedright $^\ddag$For GRB000901, the distance and energy output
were applied for an identification with EQ J084834.9-785352, see Table~\ref{table1}; $^\dag$ For GRB970402, the spectral modeling was limited to 2-10 keV, because the
spectrum shows a sharp drop above 10 keV; $^\S$ For event 010315 from GT Mus, the spectral modeling was limited to 2-13 keV, because the spectrum shows a sharp drop above 13 keV.
}

\end{document}